\documentclass[preprint,aps,nofootinbib]{revtex4-1}
\pdfoutput=1
\usepackage{bm, amsmath,amssymb,amsfonts,slashed,array,graphicx,soul}
\usepackage[vcentermath,noautoscale]{youngtab}
\usepackage{mathrsfs}
\usepackage{xcolor}
\usepackage{hyperref}
\usepackage{cancel}
\usepackage[normalem]{ulem}
\usepackage{slashed}            % for slashed characters in math mode
\usepackage{tikz}
\usetikzlibrary{arrows,shapes}
\usetikzlibrary{trees}
\usetikzlibrary{matrix,arrows} 				% For commutative diagram
\usetikzlibrary{positioning}				% For "above of=" commands
\usetikzlibrary{calc,through}				% For coordinates
\usetikzlibrary{decorations.pathreplacing}  % For curly braces
\usetikzlibrary{decorations.pathmorphing}	% For Feynman Diagrams
\usetikzlibrary{decorations.markings}
\usetikzlibrary{snakes}
\usepackage[normalem]{ulem}
\tikzset{
    vector/.style={decorate, decoration={snake}, draw},
	provector/.style={decorate, decoration={snake,amplitude=2.5pt}, draw},
	antivector/.style={decorate, decoration={snake,amplitude=-2.5pt}, draw},
    fermion/.style={draw, postaction={decorate},
        decoration={markings,mark=at position .55 with {\arrow[draw]{>}}}},
    fermionbar/.style={draw, postaction={decorate},
        decoration={markings,mark=at position .55 with {\arrow[draw=black]{<}}}},
    fermionnoarrow/.style={draw},
    gluon/.style={decorate, draw,decoration={coil,amplitude=4pt, segment length=6pt}, line width=1},
    scalar/.style={dashed,draw, postaction={decorate},
        decoration={markings,mark=at position .55 with {\arrow[draw]{>}}}},
    scalarbar/.style={dashed,draw, postaction={decorate},
        decoration={markings,mark=at position .55 with {\arrow[draw]{<}}}},
    scalarnoarrow/.style={dash pattern = on 6 pt off 3 pt,draw},
    electron/.style={draw, postaction={decorate},
        decoration={markings,mark=at position .55 with {\arrow[draw]{>}}}},
	bigvector/.style={decorate, decoration={snake,amplitude=4pt}, draw},
	vectorscalar/.style={loosely dotted,draw, postaction={decorate}},
}

\def\lsim{\mathrel{\rlap{\lower4pt\hbox{\hskip1pt$\sim$}}
    \raise1pt\hbox{$<$}}}
\def\gsim{\mathrel{\rlap{\lower4pt\hbox{\hskip1pt$\sim$}}
    \raise1pt\hbox{$>$}}}

\renewcommand{\thefootnote}{\fnsymbol{footnote}}
\begin{document}
\title{Exotic Electroweak Signals in Twin Higgs}
\author{Hsin-Chia Cheng,$^{1}$ Ennio Salvioni$^{1,2}$ and Yuhsin Tsai$^{3\;}$}
\email[Email: ]{cheng@physics.ucdavis.edu}\email{ennio.salvioni@tum.de}\email{yhtsai@umd.edu}
\affiliation{$^1$Physics Department, University of California, Davis, Davis, CA 95616, USA\\
$^{2}$Physik-Department, Technische Universit\"at M\"unchen, 85748 Garching, Germany \\
$^{3}$Maryland Center for Fundamental Physics, Department of Physics, University of Maryland, College Park, MD 20742, USA}
%\date{\today}
\begin{abstract}
The Twin Higgs model is the preeminent example of a theory of neutral naturalness, where the new particles that alleviate the little hierarchy problem are Standard Model (SM) singlets. The most promising collider search strategy, based on rare Higgs decays, is nevertheless not effective in significant regions of the parameter space of the low energy theory. This underlines the importance of phenomenological studies on ultraviolet completions of the Twin Higgs model, which must lie at a scale lower than $5\,$-$10$ TeV. We pursue this course in the context of non-supersymmetric completions, focusing on exotic fermions that carry SM electroweak and twin color charges, as well as on exotic vectors that transform as the bi-fundamental of the electroweak or color groups. Both $Z_2$-preserving and $Z_2$-breaking mass spectra are considered for the exotic fermions. In the former case they must be heavier than $\sim 1$ TeV, but can still be sizably produced in the decays of the color bi-fundamental vector. In the $Z_2$-breaking scenario, the exotic fermions can have masses in the few hundred GeV range without significantly increasing the fine-tuning. Once pair-produced through the electroweak interactions, they naturally form bound states held together by the twin color force, which subsequently annihilate back to SM particles. The associated resonance signals are discussed in detail. We also outline the phenomenology of the electroweak bi-fundamental vectors, some of which mix with the SM $W$ and $Z$ and can therefore be singly produced in hadron collisions. 
\end{abstract}
\preprint{TUM-HEP-1073-16}
\maketitle
%\tableofcontents
%%%%%%%%%%%%%%%%%%%%%%%
\renewcommand{\thefootnote}{\arabic{footnote}}
\section{Introduction}
Over the past few years, the increasingly stronger exclusion limits set by the Large Hadron Collider (LHC) on colored top partners have put significant pressure on the classic solutions to the hierarchy problem. The absence of beyond-the-Standard Model (BSM) signals has, at the same time, sparked a renewed interest in models of neutral naturalness, where the top partners do not carry Standard Model (SM) color charge. The prime example in this class is the Twin Higgs (TH) model \cite{Chacko:2005pe}, where the top and gauge partners are complete singlets under the SM, and the sensitivity to ultraviolet (UV) scales is softened thanks to a discrete $Z_2$ symmetry. 

By construction, the collider phenomenology of the lowest-lying states in TH is very challenging, and rare decays of the Higgs into long-lived twin particles typically constitute the most promising signatures \cite{Craig:2015pha,Curtin:2015fna,Csaki:2015fba}. However, these signals display a strong sensitivity to the unknown parameters of the model that limits to some extent their robustness. For example, in the Fraternal TH model \cite{Craig:2015pha}, where only the third generation of twin particles is introduced\footnote{In the Fraternal TH, the twin tau and twin neutrino are introduced to cancel gauge anomalies. Alternatively, one can consider models with vector-like twin quarks \cite{Craig:2016kue}, where the twin leptons are not necessary.} to avoid the cosmological problems associated with light degrees of freedom, the lightest twin glueball can be long-lived. It decays into SM particles through mixing with the Higgs, with a width proportional to $\Lambda^{-7}$, where $\Lambda$ is the confinement scale of twin QCD. As a consequence, the twin glueball displaced decays are observable at the LHC only in a relatively narrow range of twin confinement scales, while naturalness considerations allow a wider uncertainty on $\Lambda$. On the other hand, a robust deviation from the SM is provided by Higgs couplings modifications proportional to $v^2/f^2$ that arise due to the pseudo-Goldstone nature of the Higgs, where $f$ is the scale of spontaneous $Z_2$ breaking and $v$ is the electroweak symmetry breaking (EWSB) Higgs vacuum expectation value (VEV). Still, the future LHC precision on Higgs couplings measurements will be limited to $f\lesssim 3\,$-$\,4\,v$ \cite{Burdman:2014zta}. Altogether, these considerations suggest that it is not inconceivable that the LHC may remain blind to a TH model with $f\sim 1$ TeV. 

The low-energy theory of TH, however, requires UV completion at a relatively low scale $< 4\pi f \sim 10$ TeV. The extended theory necessarily contains new particles, some of which can give visible signals at colliders. While more model-dependent, these signatures may turn out to be key to the discovery of the model. For example, non-supersymmetric UV completions generically predict the existence of new exotic fermions, charged under both the SM and twin gauge symmetries. These vector-like fermions were already introduced in the original TH paper \cite{Chacko:2005pe}, to cut off the residual logarithmic divergences in the Higgs potential. Furthermore, they appear in composite TH completions \cite{Barbieri:2015lqa,Low:2015nqa}, where they are resonances of the strong sector, and in UV completions with extra dimensions \cite{Geller:2014kta,Craig:2014aea,Craig:2014roa}, where they are Kaluza-Klein (KK) excitations of bulk fields whose zero modes are removed through boundary conditions or orbifold projections. Some of these fermions carry SM color (as well as twin electroweak) charge and can therefore be produced with large rates at the LHC or at a future $100$ TeV collider, depending on their masses. The phenomenology of these `exotic quarks' was presented in Ref.~\cite{Cheng:2015buv}, where it was shown that future searches for their signals can test large regions of the parameter space of the Fraternal TH.\footnote{In weakly coupled UV completions, including supersymmetric ones \cite{Falkowski:2006qq,Chang:2006ra,Craig:2013fga,Katz:2016wtw}, a rather robust phenomenological feature is provided by the scalar radial mode, which is expected to be narrow and relatively light. It is produced through mixing with the SM-like Higgs and can be discovered through its decays to SM dibosons \cite{Buttazzo:2015bka}.} 

%\enlargethispage{-15pt}
In this paper we explore the phenomenology of other states that can be expected to accompany the exotic quarks (which were labeled by $\tilde{q}_3^A$ in Ref.~\cite{Cheng:2015buv}), in a non-supersymmetric UV completion of TH. We focus primarily on the mirror partners of the exotic quarks, which are vector-like `exotic fermions' that carry twin color and SM electroweak charges (labeled by $\tilde{q}_3^B$ in Ref.~\cite{Cheng:2015buv}). In addition we consider exotic vector particles, including both bifundamentals under $SU(2)_A \times SU(2)_B$, where $A$ and $B$ denote the SM and twin weak groups, respectively, and bifundamentals under the color symmetries $SU(3)_A \times SU(3)_B$. The electroweak bifundamentals, which we label $\mathcal{W}$, are necessary to restore at high energy the $SU(4)$ (or $SO(8)$) symmetry that protects the pseudo-Goldstone Higgs. On the other hand, the $SU(3)$ bifundamentals, dubbed $\mathcal{X}$, appear in models where the SM and twin color groups are embedded into an $SU(6)$ symmetry. This is not strictly required for a consistent UV completion, so the presence of $\mathcal{X}$ is somewhat more model-dependent.

As discussed in Ref.~\cite{Cheng:2015buv}, stop searches based on $t\bar{t}$ plus missing transverse energy (MET) constitute a robust probe of the exotic quarks. Given the absence of any signals in the first $\sim 15$ fb$^{-1}$ of data collected by each of ATLAS and CMS at $13$ TeV, we estimate that the current lower bound on the vector-like mass of the $\tilde{q}_3^A$ is approximately $\tilde{M}_A \gtrsim 1$ TeV. If the masses of the exotic fermions respect the $Z_2$ symmetry, as we assume in the first part of this paper, the same lower bound applies to $\tilde{M}_B$, the mass of the $\tilde{q}_3^B$. As a consequence the uncolored exotic fermions have a very suppressed electroweak pair-production, even at a $100$ TeV collider. On the other hand, if the colored exotic vector $\mathcal{X}$ is heavier than the exotic fermions, its decays can provide a sizable production rate for the $\tilde{q}_3^B$. We sketch the corresponding phenomenology, finding that it is qualitatively similar to that of the exotic quarks: the most promising signature is $bW\bar{b}W$, accompanied either by large missing transverse energy (MET) or by the displaced decay of a twin particle \cite{Cheng:2015buv}. We find that the best strategy to pin down the presence of the $\mathcal{X}$ and $\tilde{q}_3^B$ would be to require an additional $Z$ boson in the final state, since this is generated very rarely in the decays of the exotic quarks. We also outline the phenomenology of the electroweak bifundamental vectors, showing that some of them can mix with the SM $W$ and $Z$, and therefore be singly produced in the Drell-Yan process. On the other hand, we find that the $S$ parameter of electroweak precision tests (EWPT) requires their masses to be at least $\sim 3$ TeV, implying that they are likely out of the LHC reach, but may be discovered at a $100$ TeV collider. 

Notice that from a phenomenological perspective, $\tilde{M}_A$ is independent from $\tilde{M}_B$. Furthermore, $\tilde{M}_A \neq \tilde{M}_B$ breaks the $Z_2$ only softly, thus preserving the cancellation of $1$-loop quadratic divergences in the Higgs mass. Since $\tilde{q}_{3}^B$ only couples to the SM through the electroweak interactions, the experimental constraints still allow it to be relatively light. It is therefore logical to analyze the region of parameter space where $\tilde{M}_B \ll \tilde{M}_A \sim \mathrm{TeV}$, to which the second part of this paper is devoted. We find that for $\tilde{M}_B$ in the few hundred GeV range,  one of the exotic fermions, $\mathcal{K}^-$, naturally has a very suppressed decay width. This implies that once produced through the electroweak interactions, $\bar{\mathcal{K}}^+ \mathcal{K}^-$ pairs form bound states held together by the twin strong force, which eventually annihilate back to the SM. We study in detail the associated resonance signals, which provide a novel phenomenological aspect of TH models. In addition, we consider the significant effects that a relatively light color bifundamental vector may have on the bound state decays. A light $\mathcal{X}$ is expected to be accompanied by a light excited gluon (for which we will use the name `KK gluon' in analogy to an extra-dimensional model) in realistic models, so for completeness, we also summarize the main constraints on the KK gluon. Finally, we inspect closely the consequences on naturalness of the $Z_2$-breaking exotic fermion masses. While the $1$-loop effects are mild, we identify a $2$-loop quadratically divergent contribution to the Higgs mass that can be important for light $\mathcal{X}$. 

The remainder of the paper is organized as follows. In Sec.~\ref{Exoparticles} we introduce the exotic states that are the subject of this paper. For concreteness we do this in the context of a two-site model, which provides a convenient minimal description, but many of our results are general, and also apply to more elaborate constructions. Section~\ref{Z2preserve_pheno} presents the phenomenology of the scenario where the exotic fermion masses respect the $Z_2$ symmetry. In \ref{subs pheno:1} we compute the pair-production cross section of the color bifundamental vector, and estimate the main signals arising from cascade decays that involve the $\tilde{q}_3^B$. The salient properties of the weak bifundamental vectors are discussed in \ref{subs pheno:2}. Section~\ref{Z2break} contains the discussion of the $Z_2$-breaking scenario, which gives the main novel results of our paper. The phenomenology of $(\bar{\mathcal{K}}^+\mathcal{K}^-)$ bound states is studied in~\ref{subs Z2break:1}, whereas \ref{subs Z2break:2} focuses on the bound states containing the $SU(2)_A$ partner of $\mathcal{K}^-$, whose signals are more model-dependent. In \ref{subs Z2break:3} we discuss the effects of a light $\mathcal{X}$ on the bound state physics, and in \ref{subs Z2break:4} we summarize the main properties of the KK gluon, which is expected to have a mass comparable to that of $\mathcal{X}$. To conclude the section, the consequences on naturalness of the $Z_2$ breaking in the exotic fermion masses are presented in \ref{subs Z2break:5}. Finally, our conclusions are drawn in Sec.~\ref{conclusions}. For the sake of completeness, the detailed construction of a two-site model for the electroweak sector is given in App.~\ref{SU4exo}. Appendix~\ref{SO8exo} contains some details about the additional states that appear if the global symmetry of the TH is extended from $SU(4)$ to $SO(8)$, which ensures custodial protection of the $T$ parameter. The phenomenologies of the new exotic states are qualitatively similar to the ones already considered.

%%%%%%%%%%%%%%%
\section{Exotic particles in Twin Higgs models}\label{Exoparticles}
The simplest way to introduce the exotic fermions is to follow the approach of the original TH paper \cite{Chacko:2005pe}, where the symmetries of the top Yukawa were extended to $SU(6)\times SU(4) \times U(1)_X$. Adopting the notation of Ref.~\cite{Cheng:2015buv}, where in particular the SM $SU(2)_A$ is embedded in the lower right corner of $SU(4)$, we have
\begin{align} 
-\mathcal{L}_t \,=&\, y_t H^\dagger Q_{3L} \bar{u}_{3R} + \mathrm{h.c.} \nonumber \\ \,=&\, y_t \begin{pmatrix} H_B^\dagger & H_A^\dagger\end{pmatrix}\begin{pmatrix} q^B_{3L} & \tilde{q}^A_{3L} \\ \tilde{q}^B_{3L} & q^A_{3L} \end{pmatrix} \begin{pmatrix} \overline{u}^B_{3R} \\ \overline{u}^A_{3R} \end{pmatrix} + \mathrm{h.c.},
\label{eq:yukawa}
\end{align}
where $Q_{3L} \sim (\mathbf{6}, \mathbf{4}, 1/12)$, $u_{3R} \sim (\mathbf{6}, \mathbf{1}, 1/3)$ and $H\sim (\mathbf{1}, \mathbf{4}, -1/4)$. The color and twin color gauge groups are embedded in the diagonal of $SU(6)$. The `exotic' fermion doublets $\tilde{q}^A_{3}$ and $\tilde{q}^B_{3}$ are given vector-like masses,
\begin{equation}
-\mathcal{L}_{m} = \tilde{M}_A\overline{\tilde{q}}^{A}_{3R} \tilde{q}_{3L}^A + \tilde{M}_B \overline{\tilde{q}}^{B}_{3R} \tilde{q}_{3L}^B + \mathrm{h.c.}.
\label{eq:mass}
\end{equation}
If the masses in Eq.~\eqref{eq:mass} respect the $Z_2$ symmetry, then $\tilde{M}_A = \tilde{M}_B = \tilde{M}$ cuts off the logarithmic divergences in the Higgs potential arising from Eqs.~(\ref{eq:yukawa}),~(\ref{eq:mass}), leaving a finite and calculable result. In this paper we also consider the $Z_2$-breaking scenario $\tilde{M}_A \neq \tilde{M}_B$, in which case the residual logarithmic divergences need to be cut off by additional states with larger, $Z_2$-symmetric masses.

The phenomenology of the states belonging to $\tilde{q}^A_3$, which carry SM color and twin electroweak charges, was extensively discussed in Ref.~\cite{Cheng:2015buv}. Inserting in Eq.~\eqref{eq:yukawa} the expression of the Higgs field in the unitary gauge,
\begin{equation} \label{H UG}
H  = \frac{f}{\sqrt{2}}\begin{pmatrix} \cos \frac{h}{f} \\ 0 \\ \sin \frac{h}{f} \\ 0 \end{pmatrix},
\end{equation}
leads to the mass mixing of the up-type component $\tilde{u}_3^A$ with the top quark, and the corresponding heavy mass eigenstate was labeled $\mathcal{T}$. The down-type component does not mix, and was labeled $\mathcal{B}$. These exotic quarks are pair-produced through QCD and decay into SM tops plus twin gauge bosons $\hat{W},\hat{Z}$ (henceforth we denote the twin partners of the SM particles with a hat), followed by the decay of the twin gauge bosons into twin fermions. Thus the `irreducible' signal of the exotic quarks is $t\bar{t}$+MET, where the twin particles escape detection. In addition, depending on the parameters in the twin sector, some of the twin particles produced in the cascade can decay back to the SM with long lifetimes, giving rise to $t\bar{t}$+displaced vertex signatures. For example, the decay $\hat{Z} \to \bar{\hat{b}}\hat{b}$ is followed by twin hadronization, and some of the resulting twin bottomonia or twin glueballs can have macroscopic lifetimes. The twin leptons that arise from the decay of $\hat{W}$ can also give displaced signals, through mixing with the SM neutrinos. The reach of the searches for these exotic signatures extends above $2$ TeV at the LHC and above $10$ TeV at a future $100$ TeV collider, often surpassing that of the searches for stop-like signals based on large MET \cite{Cheng:2015buv}. 

Notice that the mass mixing between $\tilde{u}_3^A$ and the top quark implies the theoretical lower bound \cite{Cheng:2015buv}
\begin{equation} \label{MAtheo}
\tilde{M}_A \gtrsim \sqrt{2}\, m_t\,\frac{f}{v} \sim y_t f\,.
\end{equation}
In addition, we estimate that stop searches currently set an experimental lower bound
\begin{equation} \label{MAexp}
\tilde{M}_A \gtrsim 1\; \mathrm{TeV}.
\end{equation}
In this paper we focus instead on the fermions contained in $\tilde{q}_3^B$, which carry twin color and SM electroweak charges and whose phenomenology was so far unexplored.

In a UV completion, it is also plausible that new vector particles exist, that allow for the restoration of the full $SU(6)\times SU(4)\,[\text{or} \times SO(8)]$ symmetry at high energies. For $SU(6)$, the new states include `exotic vectors' which transform as the bifundamental representation of $SU(3)_A \times SU(3)_B$, denoted by $\cal{X}_\mu$ in this work, as well as the excited state of the gluon (labeled $\mathcal{G}_\mu$) and that of the twin gluon. For $SU(4)$, among the new vectors we expect exotics which transform as the bifundamental of $SU(2)_A\times SU(2)_B$, denoted by $\mathcal{W}_\mu$, in addition to excitations of the SM and twin gauge bosons. All these particles can be described, for example, in a two-site model. On one site, there is a full $SU(6)\times SU(4)$ gauge symmetry. All the fields in Eq.~(\ref{eq:yukawa}) live on this site, so the Yukawa interaction respects the symmetry. On the second site, only $SU(3)^2 \times SU(2)^2$ is gauged. All the light SM and twin fermions, as well as the right-handed exotic fermions $\tilde{q}_{3R}^{A,B}$, live on this second site. The gauge symmetries on the two sites are broken down to the diagonal subgroup by the VEVs of link fields, which also generate the mass terms for the exotic fermions in Eq.~\eqref{eq:mass}. For a strongly coupled UV completion, the first site can be viewed as the hidden local symmetry from the strong dynamics~\cite{Bando:1984ej}. Its gauge coupling, which we denote $g_{\mathcal{X}}$ for $SU(6)$ and $g_\rho$ for $SU(4)$, is expected to be very large and the particles living on that site are composite degrees of freedom of the strong dynamics. The second site represents the elementary gauge fields and fermions. By varying the hierarchy between the VEV $f$ of the Higgs field $H$ and $f_d$ of the $SU(4)$ link field, our two-site model interpolates between different UV realizations, along the lines of Ref.~\cite{Cheng:2006ht}. In particular, for $f_d \ll f$ it can be viewed as a deconstruction~\cite{ArkaniHamed:2001ca,Hill:2000mu} of an extra dimensional model~\cite{Craig:2014aea,Geller:2014kta}. The details of the two-site model are given in App.~\ref{SU4exo}.

Notice that $SU(4)$ does not contain the custodial symmetry that protects the $T$ parameter, which therefore would generically receive large corrections at tree level. However, this difficulty can be removed by extending the UV symmetry to $SO(8)$ \cite{Chacko:2005un}, which guarantees that $T=0$ at tree level. Since $SU(4)\subset SO(8)$, the exotic particles studied here are automatically present also in the extended model. In addition, even though the $SO(8)$ model contains additional exotics, their phenomenology is qualitatively well captured by the analysis performed in this paper, as explained in App.~\ref{SO8exo}.

\begin{table}
   \begin{center}
   \begin{tabular}{c|c c|c c|c c|c c} % Column formatting, @{} suppresses leading/trailing space
  & \multicolumn{2}{c|}{$SU(3)$} & \multicolumn{2}{c|}{$SU(2)$} & \multicolumn{2}{c|}{$U(1)$} & \multicolumn{2}{c}{$U(1)_{\rm em}$} \\
  &  $\,A$ & $B$ & $A$ & $B$ & $Y$ & $D$ & SM & Twin \\
   \hline
   &&&&&& \\[-0.7cm]    
   $\tilde{q}^A_3 = \begin{pmatrix} \tilde{u}^A_3 \approx \mathcal{T} \\ \tilde{d}^A_3 = \mathcal{B} \end{pmatrix}$  & $\;\mathbf{3}$ & $\;\mathbf{1}$ & $\;\mathbf{1}$ & $\;\mathbf{2}$ & $2/3$ & $-1/2$ & $\begin{pmatrix} 2/3 \\ 2/3 \end{pmatrix}$ & $\begin{pmatrix}    0 \\ -1 \end{pmatrix}$ \\      
   &&&&&& \\[-0.7cm]   $\tilde{q}^B_3 = \begin{pmatrix} \tilde{u}^B_3 \approx \mathcal{K}^0 \\ \tilde{d}^B_3 = \mathcal{K}^- \end{pmatrix}$  & $\;\mathbf{1}$ & $\;\mathbf{3}$ & $\;\mathbf{2}$ & $\;\mathbf{1}$ & $-1/2$ & $2/3$ & $\begin{pmatrix} 0 \\ -1 \end{pmatrix}$ &    $\begin{pmatrix} 2/3 \\ 2/3 \end{pmatrix}$ \\
   \hline
   &&&&&& \\[-0.7cm]    
   $\mathcal{X}$  & $\;\mathbf{3}$ & $\;\bar{\mathbf{3}}$ & $\;\mathbf{1}$ & $\;\mathbf{1}$ & $2/3$ & $-2/3$ & $2/3$ & $-2/3$ \\
   &&&&&& \\[-0.7cm]    
   $\mathcal{W} = \begin{pmatrix} \mathcal{W}_1^0 & \mathcal{W}_1^+ \\ \mathcal{W}_2^0 & \mathcal{W}_2^+ \end{pmatrix}$  & $\;\mathbf{1}$ & $\;\mathbf{1}$ & $\;\mathbf{2}$ & $\;\mathbf{2}$ & $1/2$ & $-1/2$ & $\begin{pmatrix} 0 & 1 \\ 0 &    
   1\end{pmatrix}$  & $\begin{pmatrix} 0 & 0 \\ -1 & -1\end{pmatrix}$ \\      
   \end{tabular}   
%   }
   \end{center}
   \caption{Quantum numbers of the exotic fields under the SM and twin gauge symmetries. ($U(1)_D$ is the twin hypercharge.) The fields in the upper part of the table are Dirac fermions, while those in the lower part are complex vectors.}
   \label{Tab:charges}
\end{table}

\section{$Z_2$ - preserving phenomenology}\label{Z2preserve_pheno}
In this section we outline the phenomenology of the exotic off-diagonal states, with the exception of the exotic quarks $\tilde{q}_{3}^A$, which were thoroughly studied in Ref.~\cite{Cheng:2015buv}. The quantum numbers of the particles are collected in Table~\ref{Tab:charges}. In this section we assume that the masses of the exotic fermions respect the $Z_2$ symmetry.
\subsection{Exotic fermions and $SU(6)$ vectors} \label{subs pheno:1}
In addition to the mixing of $\tilde{u}_3^A$ with the SM top quark, the Lagrangian in Eqs.~(\ref{eq:yukawa}),~(\ref{eq:mass}) yields a mixing of the top-component exotic fermion $\tilde{u}_3^B$ with the twin top, 
\begin{equation} \label{massmatrix}
-\begin{pmatrix} \bar{u}_{3R}^B & \bar{\tilde{u}}_{3R}^B \end{pmatrix} \begin{pmatrix} \frac{y_t f}{\sqrt{2}}c_h & \frac{y_t f}{\sqrt{2}}s_h \\ 0 & \tilde{M}_B \end{pmatrix} \begin{pmatrix} u_{3L}^B \\ \tilde{u}_{3L}^B \end{pmatrix} + \mathrm{h.c.}\,
\end{equation}
where $s_h \equiv \sin (h/f), c_h \equiv \cos(h/f)$. The mixing is diagonalized by the rotations
\begin{equation}
\begin{pmatrix} u_{3L,R}^B \\ \tilde{u}_{3L,R}^B \end{pmatrix} \to \begin{pmatrix} -c_{L,R} & s_{L,R} \\ s_{L,R} & c_{L,R} \end{pmatrix} \begin{pmatrix} \hat{t}_{L,R}  \\ \mathcal{K}^0_{L,R}  \end{pmatrix},
\end{equation}
where $\hat{t}$ and $\mathcal{K}^0$ are mass eigenstates. For the remainder of this section we assume $\tilde{M}_A = \tilde{M}_B = \tilde{M} > y_t f/\sqrt{2}$, so $m_{\hat{t}} < m_{\mathcal{K}^0}\,$. The mixing angles are given by
\begin{equation}
s_L = \frac{m_{\hat{t}}}{\tilde{M}}s_R\,,\qquad s_R \simeq \frac{\tilde{M} y_t v/\sqrt{2}}{\tilde{M}^2 - y_t^2 f^2/2} + O(v^3),\,
\end{equation}
while the masses read, at first order in $v^2$,
\begin{equation}
m_{\hat{t}}^2 \simeq \frac{y_t^2 f^2}{2} \left(1- \frac{v^2}{f^2}\frac{\tilde{M}^2}{\tilde{M}^2 - y_t^2 f^2/2}\right)\,,\qquad m_{\mathcal{K}^0}^2 \simeq \tilde{M}^2 \left(1 + \frac{y_t^2 v^2/2}{\tilde{M}^2 - y_t^2 f^2/2}\right)\,.
\end{equation}
On the other hand, $\mathcal{K}^- \equiv \tilde{d}_{3}^B$ does not mix with any other state. 

The decays of the $\mathcal{K} \equiv \{\mathcal{K}^0,\mathcal{K}^-\}$ can be understood using the Goldstone equivalence theorem. Plugging the expression of $H$ expanded to $O(1/f)$ \cite{Cheng:2015buv} into the top Yukawa, we find, for $\tilde{M} \gg y_t f/\sqrt{2}\,$,
\begin{equation} \label{exodecays}
\mathcal{L}_t \;\;\ni\;\; -\, y_t \left[\frac{1}{\sqrt{2}} \left(i \pi_3 - h + i \frac{v}{2f}\pi_7\right) \bar{\hat{t}}_R \mathcal{K}^0_L + i \pi^+ \bar{\hat{t}}_R \mathcal{K}^-_L \right] + \mathrm{h.c.},
\end{equation}
where $\pi^+ \equiv (\pi_1 + i \pi_2)/\sqrt{2}$ and $\pi_3$ are the would-be Goldstone bosons eaten by the SM $W$ and $Z$, $h$ is the physical Higgs boson, and $\pi_7$ corresponds to the longitudinal component of the twin $\hat{Z}$. Equation \eqref{exodecays} shows that the largest decay widths of $\mathcal{K}^0$ are those into $\hat{t}Z,\hat{t}h$, with a small component of $\mathcal{K}^0 \to \hat{t}\hat{Z}$ parametrically suppressed by $v^2/f^2$.\footnote{For smaller $\tilde{M}$ the couplings mediating the additional decays $\mathcal{K}^0 \to \hat{b} \hat{W}$ and $\mathcal{K}^0 \to \mathcal{K}^- W$, which are $\propto s_R$, cannot be neglected. Notice, however, that since $m_{\mathcal{K}^0} - m_{\mathcal{K}^-} < m_W$, the latter is actually a $3$-body decay.} The $\mathcal{K}^-$, on the other hand, decays to $\hat{t} W$ with unity branching ratio. 

\begin{figure}
\begin{center}
\includegraphics[width=8cm]{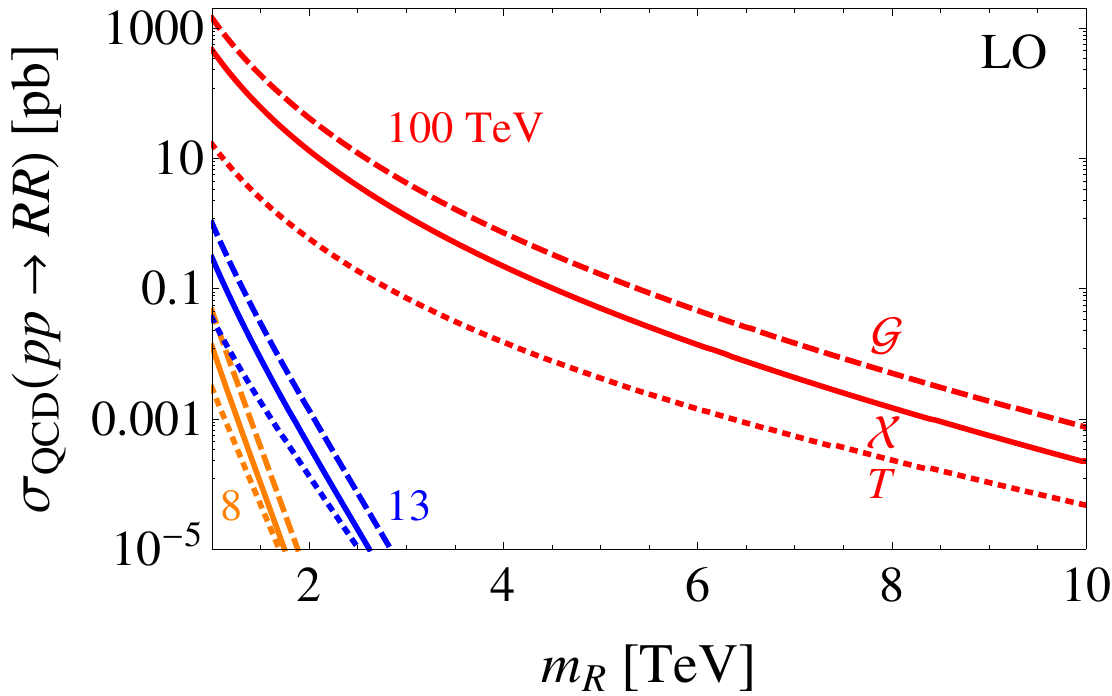}
\caption{QCD pair production cross sections at $pp$ colliders for the exotic vector $\mathcal{X}$ and KK gluon $\mathcal{G}$. In addition, for comparison we show the cross section for a Dirac fermion $T$ in the fundamental representation of color. All cross sections were computed at leading order (LO) with factorization and renormalization scales set to $\mu_{\rm fact} = \mu_{\rm ren} = \sqrt{\hat{s}}/2$, using MSTW08LO parton distribution functions (PDFs). The ratio $\sigma(\mathcal{G}\mathcal{G})/\sigma(\mathcal{X}\mathcal{X}^\ast)$ varies between 3.0 and 3.7 for the masses and energies considered. Notice that $\sigma(\mathcal{X}\mathcal{X}^\ast)$ contains the factor $N_c = 3$ resulting from the sum over twin color. The different scaling of $\sigma(T\bar{T})$ with the mass is due to the $q\bar{q}$-initiated component.}
\label{fig:RRxsec}
\end{center}
\end{figure}

The theoretical and experimental constraints in Eqs.~\eqref{MAtheo},~\eqref{MAexp} require $\tilde{M}_A \gtrsim 1$ TeV (we take $f = 1$ TeV as benchmark in this paper). For a $Z_2$-symmetric spectrum this constraint also applies to $\tilde{M}_B$, implying that the pair production of the $\mathcal{K}$ via the electroweak interactions is suppressed even at a future $100$ TeV collider. However, if $m_{\mathcal{X}}> \tilde{M}$ the decays of the exotic vector $\mathcal{X}$ can provide a much larger production rate for the uncolored exotic fermions. The $\mathcal{X}$ is a bifundamental of $SU(3)_A \times SU(3)_B$, therefore it can be pair-produced in the process $gg\to \mathcal{X}\mathcal{X}^\ast$. The couplings of $\mathcal{X}$ to the gluons arise from the kinetic Lagrangian
\begin{equation}
-\, (D_\mu \mathcal{X}_\nu)^\dagger (D^\mu \mathcal{X}^\nu - D^\nu \mathcal{X}^\mu) + i g_s G^{\mu\nu\,a}\mathcal{X}_\nu^\ast t^a \mathcal{X}_\mu, \label{exovector-g}
\end{equation}
where the covariant derivative is defined as $D_\mu \mathcal{X}_\nu = \partial_\mu \mathcal{X}_\nu - i g_s G_\mu^a t^a \mathcal{X}_\nu$. Here $t^a$, $a = 1,\ldots,8$ are the generators in the fundamental of $SU(3)_A$, and twin color indices and interactions are understood but suppressed. Notice that the coefficient of the last term in Eq.~\eqref{exovector-g} would be arbitrary, if we were only imposing the low-energy $SU(3)_A$ gauge symmetry. We have set this coefficient to the value that corresponds to $\mathcal{X}$ gauging part of the (spontaneously broken) extended $SU(6)$ gauge symmetry in the UV, as in the two-site model. Using the couplings derived from Eq.~\eqref{exovector-g}, the cross section for pair production of $\mathcal{X}$ can be readily computed
\begin{equation}
\hat{\sigma}(gg\to \mathcal{X}\mathcal{X}^\ast) \,=\, \frac{\pi \alpha_s^2}{16 \hat{s}^3} \Bigg[\beta \hat{s} \left(32\frac{\hat{s}^2}{m_\mathcal{X}^2} + 87 \hat{s} + 186\, m_\mathcal{X}^2\right) - 24 (3 \hat{s}^2 + 4 \hat{s}\, m_\mathcal{X}^2 + m_\mathcal{X}^4) \log \frac{1+\beta}{1-\beta}\Bigg], \label{gg to XX}
\end{equation}
where $\beta \equiv \sqrt{1- 4m_\mathcal{X}^2/\hat{s}}$ and an extra factor of $N_c = 3$ was included, due to the twin color sum. By convoluting with the $gg$ parton luminosity, we obtain the hadronic cross section shown in Fig.~\ref{fig:RRxsec}. 

\begin{figure}
\begin{center}
\includegraphics[width=12cm]{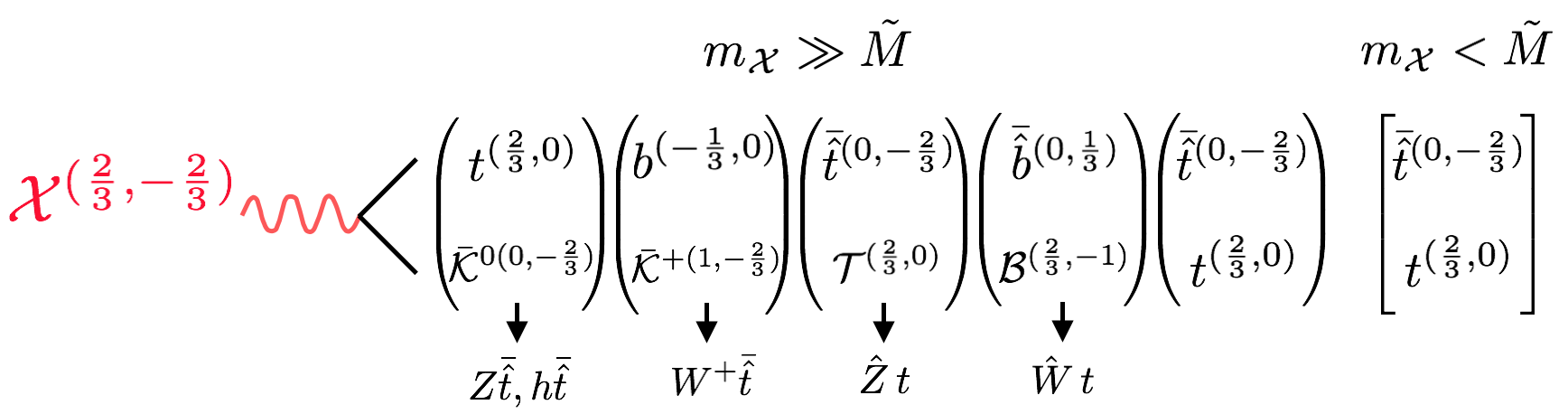}

\vspace{0.5cm}
\includegraphics[width=10cm]{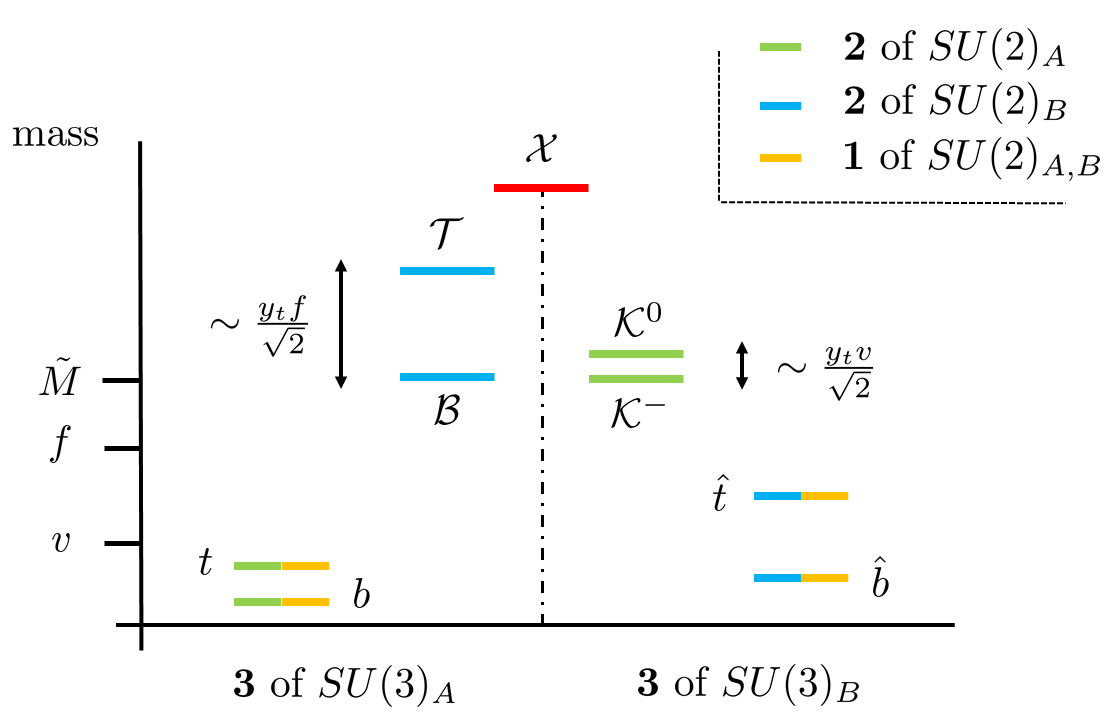}
\caption{{\it Top panel:} main decay channels of the $SU(6)$ exotic vector $\mathcal{X}$. The superscripts indicate the (SM,~twin) electric charges. {\it Bottom panel:} sketch of the colored spectrum in the $Z_2$-preserving case $\tilde{M}_{A} = \tilde{M}_{B} = \tilde{M}$, under the assumption that $m_{\cal X} > \tilde{M}$. The small $t\,$-$\mathcal{T}$ and $\hat{t}\,$-$\,\mathcal{K}^0$ mixings are neglected in the color scheme. The mass splittings among the exotic fermions are $\sqrt{m_{\mathcal{T}}^2 - m_{\mathcal{B}}^2} \sim y_t f/\sqrt{2}$ and $\sqrt{m_{\mathcal{K}^0}^2 - m_{\mathcal{K}^-}^2} \sim y_t v/\sqrt{2}\,$.}
\label{fig:spectrum_Z2sym}
\end{center}
\end{figure}

The $\cal{X}$ also couples, with strength $g_{\mathcal{X}}$, to the fermions transforming in the fundamental of $SU(6)$, namely $(\tilde{q}_{3L}^B\;q_{3L}^A)^T$, $(q_{3L}^B\;\tilde{q}_{3L}^A)^T$ and $(u_{3R}^B\;u_{3R}^A)^T$. For example,
\begin{equation} \label{exovector coupl}
\frac{g_\mathcal{X}}{\sqrt{2}}\,\bar{q}_{3L}^A \gamma^\mu \tilde{q}_{3L}^B \cal{X}_\mu + \mathrm{h.c.}.
\end{equation}       
Notice that the exotic vector carries SM electric charge $\pm 2/3$, and twin electric charge $\mp 2/3$. The right-handed bottom quarks $d_{3R}^{A,B}$ and exotic fermions $\tilde{q}_{3R}^{A,B}$, as well as the fermions of the first two generations, live instead on the $SU(3)^2$ site and therefore do not couple directly to the exotic vector. Equation \eqref{exovector coupl} dictates the decay of the $\mathcal{X}$: the main channels are $\mathcal{X} \to \bar{\hat{t}}t, \bar{\hat{t}}\mathcal{T}, \bar{\hat{b}} \mathcal{B}, \bar{\mathcal{K}}^0 t, \bar{\mathcal{K}}^+ b$, shown in the top panel of Fig.~\ref{fig:spectrum_Z2sym}. In addition, fermion mixings generate a small width for $\mathcal{X} \to \bar{ \mathcal{K}}^0\mathcal{T}$.

If $m_\mathcal{X} > \tilde{M}$, the QCD pair production of $\mathcal{X}$ followed by the decays $\mathcal{X} \to \bar{\mathcal{K}}^0 t, \bar{\mathcal{K}}^+ b$ provides the largest production mechanism for the $\mathcal{K}$. A sketch of the corresponding spectrum is given in the bottom panel of Fig.~\ref{fig:spectrum_Z2sym}. On the other hand, if $m_{\mathcal{X}}<\tilde{M}$ the only kinematically allowed decay of the exotic vector is $\mathcal{X} \to \bar{\hat{t}}t$. Under the minimal assumption that the $\hat{W}$ and twin bottom escape undetected, most decay patterns of $\mathcal{X}\mathcal{X}^\ast$ lead to the $bW\bar{b}W$+MET final state, resulting in a `stop-like' signal. Therefore it would be difficult to distinguish the $\mathcal{X}$ signal from those of the exotic quarks $\mathcal{T}$ and $\mathcal{B}$, which mostly produce $t\bar{t}$+MET final states. In addition, it would be challenging to tell the very existence of the exotic fermions $\mathcal{K}$. For this purpose the most promising channel seems to be $\mathcal{X} \to \bar{\mathcal{K}}^0 t$ followed by $\bar{\mathcal{K}}^0 \to \bar{\hat{t}}h, \bar{\hat{t}}Z$. These decays yield an extra $Z$ or $h$, resulting in signals with extra jets ($b$-tagged or not) and/or leptons. Among these the $t\bar{t}Z$+jets+MET signature is particularly interesting, because it can arise from the exotic quarks only through the suppressed $\mathcal{T}\to tZ$ decay \cite{Cheng:2015buv}. 

\subsection{Exotic $SU(4)$ vectors}\label{subs pheno:2}
The off-diagonal vectors contained in $SU(4)$ form a bidoublet of $SU(2)_A \times SU(2)_B$, which under the SM electroweak symmetry decomposes into two $\mathbf{2}_{1/2}$, $\mathcal{W}_1 = (\mathcal{W}_1^+,\,\mathcal{W}_1^0)^T$ and $\mathcal{W}_2 = (\mathcal{W}_2^+,\,\mathcal{W}_2^0)^T$. These particles can be described in a two-site model, where the gauge symmetry on the `strong' site is $SU(4)\times U(1)_X$, and on the `elementary' site it is $[SU(2)\times U(1)]^2$. The symmetries are broken to the diagonal subgroup by the VEV of a bifundamental link field $\Sigma$, $\left\langle \Sigma \right\rangle = f_d \mathbf{1}_4$, whereas the Higgs $H$ in Eq.~\eqref{H UG} transforms in the fundamental representation of the strong gauge symmetry. This leads to the following masses for the $SU(4)$ exotic vectors (see App.~\ref{SU4exo} for details)
\begin{equation} \label{exomasses}
m^2_{\mathcal{W}_{1}} = \frac{g_\rho^2}{4}(f_d^2 + f^2)   \,,\qquad m^2_{\mathcal{W}_{2}} = \frac{g_\rho^2}{4}f_d^2 \,,  
\end{equation}
where $g_\rho$ is the gauge coupling on the strong site, and the small corrections due to EWSB were neglected. An experimental lower bound on the masses of the $\mathcal{W}$ comes from the $S$ parameter of EWPT, for which the two-site structure leads to the result
\begin{equation} \label{Sformula}
\hat{S} = \frac{g^2}{g_\rho^2}\, \left(\frac{v^2}{f_d^2} + \frac{v^2}{f^2 + f_d^2}\right).
\end{equation}
Requiring $\hat{S}< 2\times 10^{-3}$ implies, choosing for example our benchmark $f = 1$ TeV and $g_\rho \sim 5$, a lower bound $f_d \gtrsim 0.85$ GeV, or $m_{\mathcal{W}_{1}} \gtrsim 3.3$ TeV and $m_{\mathcal{W}_{2}} \gtrsim 2.1$ TeV. These can be taken as rough reference for the phenomenology. As we already remarked, $SU(4)$ does not contain the custodial symmetry that protects the $T$ parameter. Therefore, tree-level corrections to $T$ would push the scale of the vector resonances much higher: as discussed in App.~\ref{SU4exo}, we find $\hat{T} \simeq v^2/f_d^2$, hence requiring that $\hat{T}\lesssim 10^{-3}$ leads to $f_d \gtrsim 8$ TeV. However, this issue can be solved by extending the UV symmetry to $SO(8) \supset SU(4)$ \cite{Chacko:2005un}, which guarantees that $T=0$ at tree level. In this case additional states are present both in the gauge and in the fermion sector, to fill multiplets of the larger symmetry. While the new states include some exotic ones, their phenomenology is not expected to be qualitatively different from that discussed here, as outlined in App.~\ref{SO8exo}.

\begin{figure}
\begin{center}
\includegraphics[width=16.cm]{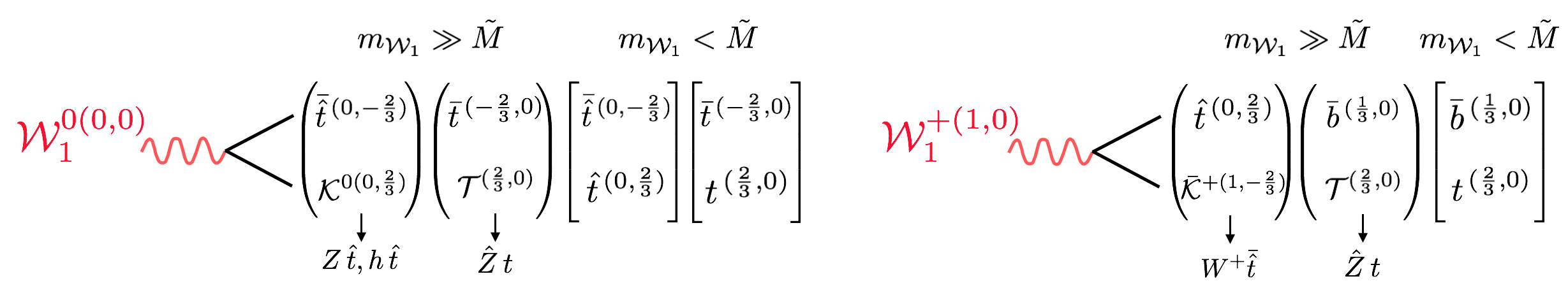}
\caption{Main decay channels of the exotic vectors contained in $\mathcal{W}_1$. The superscripts indicate the (SM,~twin) electric charges.}
\label{fig:SU4decay}
\end{center}
\end{figure}

The real component of $\mathcal{W}_1^0$, which we dub $\mathcal{W}_{1R}^0 \equiv (\mathcal{W}_1^0 + \mathrm{h.c.})/\sqrt{2}$, and $\mathcal{W}_1^+$ have the right quantum numbers to mix with the SM $Z$ and $W$, respectively, and therefore can be singly produced at hadron colliders in $\bar{q}q^{(\prime)}$ annihilation. Their decays are mediated by the couplings of the $\mathcal{W}$ to fermions, which arise from the kinetic term of $Q_{3L}\sim (\mathbf{6},\mathbf{4},1/12)$ under the $SU(6)\times SU(4) \times U(1)_X$ symmetry on the strong site. Then
\begin{equation}
i\mathrm{Tr}[\overline{Q}_{3L}\gamma^\mu D_\mu Q_{3L}] \quad \to \quad \frac{g_\rho}{\sqrt{2}}(\bar{q}_{3L}^A \slashed{\mathcal{W}}^\dagger \tilde{q}_{3L}^A + \bar{q}_{3L}^B \slashed{\mathcal{W}} \tilde{q}_{3L}^B) + \mathrm{h.c.}.
\end{equation}
Expanding in components, retaining only couplings of $\mathcal{W}_1^0$ and $\mathcal{W}_1^+$ and rotating to the basis of mass eigenstate fermions, we arrive at
\begin{align}
\frac{g_\rho}{\sqrt{2}}\Big[(s_L^{(A)} \bar{t}_L + c_L^{(A)} \overline{\mathcal{T}}_L)&\slashed{\mathcal{W}}_1^0(-c_L^{(A)} t_L + s_L^{(A)} \mathcal{T}_L) + (-c_L \bar{\hat{t}}_L + s_L \bar{\mathcal{K}}^0_L) \slashed{\mathcal{W}}_1^0 (s_L \hat{t}_L + c_L \mathcal{K}^0_L) \nonumber \\
&+ (s_L^{(A)} \bar{t}_L + c_L^{(A)} \overline{\mathcal{T}}_L)\slashed{\mathcal{W}}_1^+ b_L + (-c_L \bar{\hat{t}}_L + s_L \bar{\mathcal{K}}^0_L)\slashed{\mathcal{W}}_1^+ \mathcal{K}^-_L\Big] + \mathrm{h.c.},
\end{align}
where $s_L^{(A)}$ and $c_L^{(A)}$ are the sine and cosine, respectively, of the mixing angle between the top and the exotic quark $\mathcal{T}$ \cite{Cheng:2015buv}. Depending on the relative hierarchy between the masses of the exotic particles, we can envisage two situations:
\begin{itemize}
\item For $m_{\mathcal{W}_{1}} \gg \tilde{M}$, the dominant decays are $\mathcal{W}_1^0 \to \bar{\hat{t}}\mathcal{K}^0, \bar{t} \mathcal{T}$ and $\mathcal{W}_1^+ \to \bar{\mathcal{K}}^+\hat{t}, \bar{b}\mathcal{T}$. It follows that single production of $\mathcal{W}_1^0$ leads to the $Z,h$+MET and $\bar{t}t$+MET final states, whereas single production of $\mathcal{W}_1^+$ leads to $W^+$+MET and $\bar{b}t$+MET;
\item For $m_{\mathcal{W}_{1}} < \tilde{M}$ only the direct decays into light states are open, $\mathcal{W}_1^0 \to \bar{t}t,\bar{\hat{t}}\hat{t}$ and $\mathcal{W}_1^+ \to \bar{b}t$.
\end{itemize}
The main decays are summarized in Fig.~\ref{fig:SU4decay}. 
The lower bound $m_{\mathcal{W}_{1}}\gtrsim 3$ TeV from the $S$ parameter suggests that these exotic vectors are likely out of the LHC reach, but may be discovered at a future $100$ TeV collider. On the other hand, the states contained in $\mathcal{W}_2$ carry a non-vanishing twin electric charge, therefore their mixing with the SM gauge bosons is proportional to the breaking of the twin $U(1)_{\rm em}$, which is more model-dependent. If this breaking is negligible or absent, $\mathcal{W}_2$ can only be pair-produced through the electroweak interactions, with very suppressed cross section. Notice that in Ref.~\cite{Barbieri:2015lqa} the limit $f \gg f_d$ was considered,\footnote{Neglecting EWSB, our $f_d$ corresponds to $f$ in the notation of Sec.~2.1 of Ref.~\cite{Barbieri:2015lqa}.} in which case $\mathcal{W}_{1}$ decouples and only $\mathcal{W}_{2}$ remains in the low-energy spectrum. 

Before concluding this section, we wish to comment on the assumption we have made so far, that the twin particles produced in the cascade decays of the exotic vectors do not give any signals in the LHC detectors, effectively contributing to the missing energy. In the Fraternal TH scenario, in each event a $\bar{\hat{b}}\hat{b}$ pair is produced, held together by a string of the twin strong force. This bound state hadronizes either via string fragmentation, leading to production of twin bottomonia, or via twin gluon emission, leading to production of twin glueballs. Depending on the values of $m_{\hat{b}}$ and $\Lambda$, the resulting twin hadrons can travel a macroscopic distance before decaying within the detector, giving rise to displaced signatures. The twin leptons can also be produced through the decays of $\hat{W}$, and have long-lived decays through mixing with the SM neutrinos. In these cases, the searches for long-lived particles in association with SM objects provide a sensitivity complementary to (and possibly stronger than) that of searches based on large missing energy, as thoroughly studied for the exotic quarks in Ref.~\cite{Cheng:2015buv}.

\section{$Z_2$ - breaking scenario} \label{Z2break}
\begin{figure}
\begin{center}
\includegraphics[width=10cm]{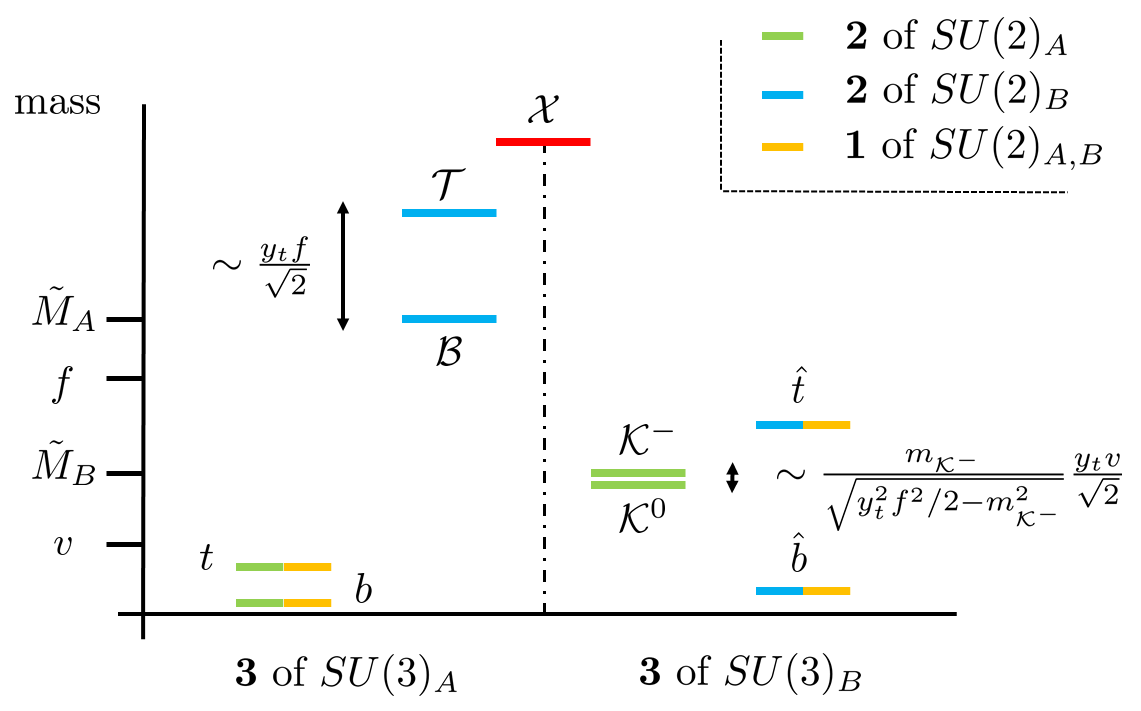}
\caption{Sketch of the colored spectrum in the $Z_2$-breaking case $\tilde{M}_{A} \gg \tilde{M}_{B}$. For definiteness we took $m_{\cal X} \gg \tilde{M}_A$. The small $t\,$-$\mathcal{T}$ and $\hat{t}\,$-$\,\mathcal{K}^0$ mixings are neglected in the color scheme. The mass splittings among the exotic fermions are $\sqrt{m_{\mathcal{T}}^2 - m_{\mathcal{B}}^2} \sim y_t f/\sqrt{2}$ and $\sqrt{m_{\mathcal{K}^-}^2 - m_{\mathcal{K}^0}^2} \sim (m_{\mathcal{K}^-} /\sqrt{y_t^2 f^2/2 - m_{\mathcal{K}^-}^2}\,)\,y_t v/\sqrt{2})\,$.
}
\label{fig:spectrum_Z2break}
\end{center}
\end{figure}

In this section we consider a region of parameters where the vector-like masses of the exotic fermions softly break the $Z_2$ symmetry, $\tilde{M}_B \ll \tilde{M}_A \sim \mathrm{TeV}$. For small $\tilde{M}_B$ the mass matrix in Eq.~\eqref{massmatrix} is diagonalized by
\begin{equation}
\begin{pmatrix} u_{3L,R}^B \\ \tilde{u}_{3L,R}^B \end{pmatrix} \to \begin{pmatrix} -c_{L,R} & s_{L,R} \\ s_{L,R} & c_{L,R} \end{pmatrix} \begin{pmatrix} \mathcal{K}^0_{L,R} \\ \hat{t}_{L,R}  \end{pmatrix}.
\end{equation}
Here $\mathcal{K}^0$ and $\hat{t}$ are mass eigenstates with $m_{\mathcal{K}^0} < m_{\hat{t}}$, where we have assumed $\tilde{M}_B < y_t f/\sqrt{2}$. For example, for $\tilde{M}_B = 350$ GeV we find $m_{\mathcal{K}^0} \simeq 336$ GeV, $m_{\mathcal{K}^-} = 350$ GeV and $m_{\hat{t}}\simeq 714$ GeV, with mixing angles $s_L \simeq 0.95$ and $s_R \simeq 0.99$. A sketch of the spectrum is given in Fig.~\ref{fig:spectrum_Z2break}.

Since $m_{\mathcal{K}^0} \lesssim m_{\mathcal{K}^-}$, the dominant decay of $\mathcal{K}^-$ is into the three-body final state $(W^\ast \to \bar{f} f')\, \mathcal{K}^0$. The partial width can be obtained by adapting the results for the decay of a chargino into a neutralino and a pair of SM fermions in a supersymmetric Standard Model~\cite{Djouadi:2001fa}. Summing over all $\bar{f}f'$ pairs, we find
\begin{equation}  \label{lambda lifetime}
\Gamma_{\mathcal{K}^-} \simeq \frac{3 G_F^2 m_{\mathcal{K}^-}^5}{160\pi^3}\left(1 - \frac{m_{\mathcal{K}^0}^2}{m_{\mathcal{K}^-}^2}\right)^5,
\end{equation}
where we neglected the small corrections due to the mixing angles, setting $s_L = s_R = 1$. We find that for the parameter region $\tilde{M}_B < 600$ GeV that we are interested in, $\Gamma_{\mathcal{K}^-} \ll \Lambda \sim \mathrm{GeV}$ is always satisfied. As a consequence, when a $\bar{\mathcal{K}}^+ \mathcal{K}^-$ pair is produced at the LHC via the electroweak interactions, it forms a bound state held together by the twin color force. Furthermore, as will be discussed in detail momentarily, the annihilation of the $(\bar{\mathcal{K}}^+ \mathcal{K}^-)$ bound states is much faster than the decay of the individual $\mathcal{K}^-$'s, leading to resonance signals. On the other hand, $\mathcal{K}^0$ decays into $\hat{W}\hat{b}$, where the twin $\hat{W}$ can be on-shell or off-shell. The corresponding width, being suppressed by $c_L^2$, satisfies $\Gamma_{\mathcal{K}^0} < \Lambda$ across all parameter space, implying the formation of additional bound states containing $\mathcal{K}^0$. However, if the $\hat{W}\hat{b}$ final state is on-shell, which occurs in a significant portion of parameter space,\footnote{Recall that for $f = 1$ TeV we have $m_{\hat{W}} \simeq 314$ GeV.} the decay of the individual $\mathcal{K}^0$'s is faster than the annihilation of $(\bar{\mathcal{K}}^0 \mathcal{K}^0)$ and $(\bar{\mathcal{K}}^0\mathcal{K}^-)$ bound states. Thus we will first concentrate on the signals from $(\bar{\mathcal{K}}^+\mathcal{K}^-)$ bound states, which appear to be more generic, and comment later about the observability of the $(\bar{\mathcal{K}}^0\mathcal{K}^0)$ and $(\bar{\mathcal{K}}^0\mathcal{K}^-)$ bound states, which requires the $\mathcal{K}^0$ to have a $3$-body decay, \emph{i.e.} $m_{\mathcal{K}^0} \lesssim m_{\hat{W}} + m_{\hat{b}}$. 

\subsection{$(\bar{\mathcal{K}}^+ \mathcal{K}^-)$ bound states}\label{subs Z2break:1}
\begin{figure}
\begin{center}
\includegraphics[width=13.0cm]{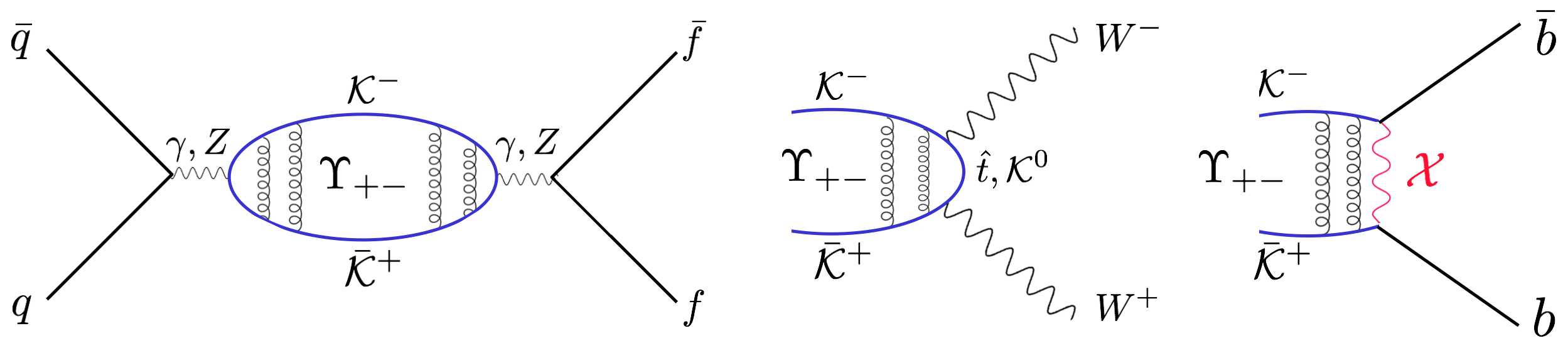}
\caption{Diagrams mediating the production and two-body decays of $\Upsilon_{+-}\,$. In the middle diagram the $W$'s are longitudinally polarized, and the amplitude is dominated by exchange of a virtual twin top.}
\label{fig:Upsilon decay}
\end{center}
\end{figure}

The $s$-wave $(\bar{\mathcal{K}}^+\mathcal{K}^-)$ bound states are a pseudoscalar and a vector, which we label $\eta_{+-}$ and $\Upsilon_{+-}$, respectively, following the SM bottomonium conventions. At hadron colliders, the $\Upsilon_{+-}$ is produced in $q\bar{q}$ annihilation via $s$-channel $\gamma/Z$, see the first diagram of Fig.~\ref{fig:Upsilon decay}. The cross section is proportional to the width for the decay $\Upsilon_{+-} \to q\bar{q}$,\footnote{The couplings of the fermion $\psi = \{\mathcal{K}^-,q\}$ to the photon and $Z$ are defined as $e Q_\psi \bar{\psi}\gamma^\mu \psi A_\mu$ and $e/(s_w c_w) \bar{\psi}\gamma^\mu(V_\psi + A_\psi) \psi Z_\mu$, respectively.}
\begin{align}
\Gamma^{\Upsilon_{+-}}_{q\bar{q}} \,=&\, N_c^2 \frac{4\pi \alpha^2}{3}\frac{\left|\psi(0)\right|^2}{m_{\mathcal{K}^-}^2}\nonumber \\
\,\times &\,\left[\Bigg(Q_{\mathcal{K}^-} Q_q + \frac{s_w^{-2} c_w^{-2} V_{\mathcal{K}^-} V_q}{1-\frac{m_Z^2}{4m_{\mathcal{K}^-}^2}}\Bigg)^2 + \Bigg(\frac{s_w^{-2} c_w^{-2} V_{\mathcal{K}^-} A_q}{1-\frac{m_Z^2}{4m_{\mathcal{K}^-}^2}}\Bigg)^2 \right],
\end{align}
where $\psi(0)$ is the wavefunction at the origin. We will return momentarily to the evaluation of this quantity. The cross section for Drell-Yan (DY) production is then
\begin{equation} 
\sigma_{q\bar{q}}(pp \to \Upsilon_{+-}) \,=\, \frac{12\pi^2}{N_c^2} \sum_{q = u,d} \frac{\Gamma^{\Upsilon_{+-}}_{q\bar{q}}}{2 m_{\mathcal{K}^-}}\frac{L_{q\bar{q}}(\frac{4m_{\mathcal{K}^-}^2}{s})}{s}  \label{drellyan}
\end{equation}
where $L_{q\bar{q}}(\tau) = \int_\tau^1 \frac{dx}{x}\left[q(x)\bar{q}(\tau/x) + q(\tau/x)\bar{q}(x)\right]$ is the $q\bar{q}$ parton luminosity. The parton luminosities are computed using the MSTW08NLO \cite{Martin:2009iq} PDFs evaluated at $\mu_{\rm fact} = m_{\mathcal{K}^-}$.

Since $\tilde{M}_B \gg \Lambda \sim O(1$-$10)$ GeV, to evaluate $\psi(0)$ we apply the standard Coulomb approximation, where the effects of confinement are neglected. Then for the ground state (see for example Ref.~\cite{Kats:2012ym} for a general discussion)
\begin{equation} \label{wavefunctionCoulomb}
\frac{\left|\psi(0)\right|^2}{m_{\mathcal{K}^-}^3} = \frac{C_F^3 \hat{\alpha}^3_s(q_{\rm rms})}{8\pi}\,,
\end{equation}
where $C_F = 4/3$ is the quadratic Casimir of the fundamental representation of $SU(3)$, and $\hat{\alpha}_s(q_{\rm rms})$ is the running twin QCD coupling evaluated at $q_{\rm rms}$, the inverse of the average distance between the two constituents, related to the Bohr radius $a_0 = 2/[C_F \hat{\alpha}_s (q_{\rm rms})m_{\mathcal{K}^-}]$ by $q_{\rm rms} = (\sqrt{3}\,a_0)^{-1}$. Using Eq.~\eqref{wavefunctionCoulomb} we compute the $\Upsilon_{+-}$ production cross section at the LHC: for example, for $\tilde{M}_B = 350$ GeV, $\sigma_{q\bar{q}}(pp \to \Upsilon_{+-})$ at $13$ TeV varies between $3.9$ and $31$ fb for $\Lambda \in [1,10]$ GeV.

\begin{figure}
\begin{center}
\includegraphics[width=8cm]{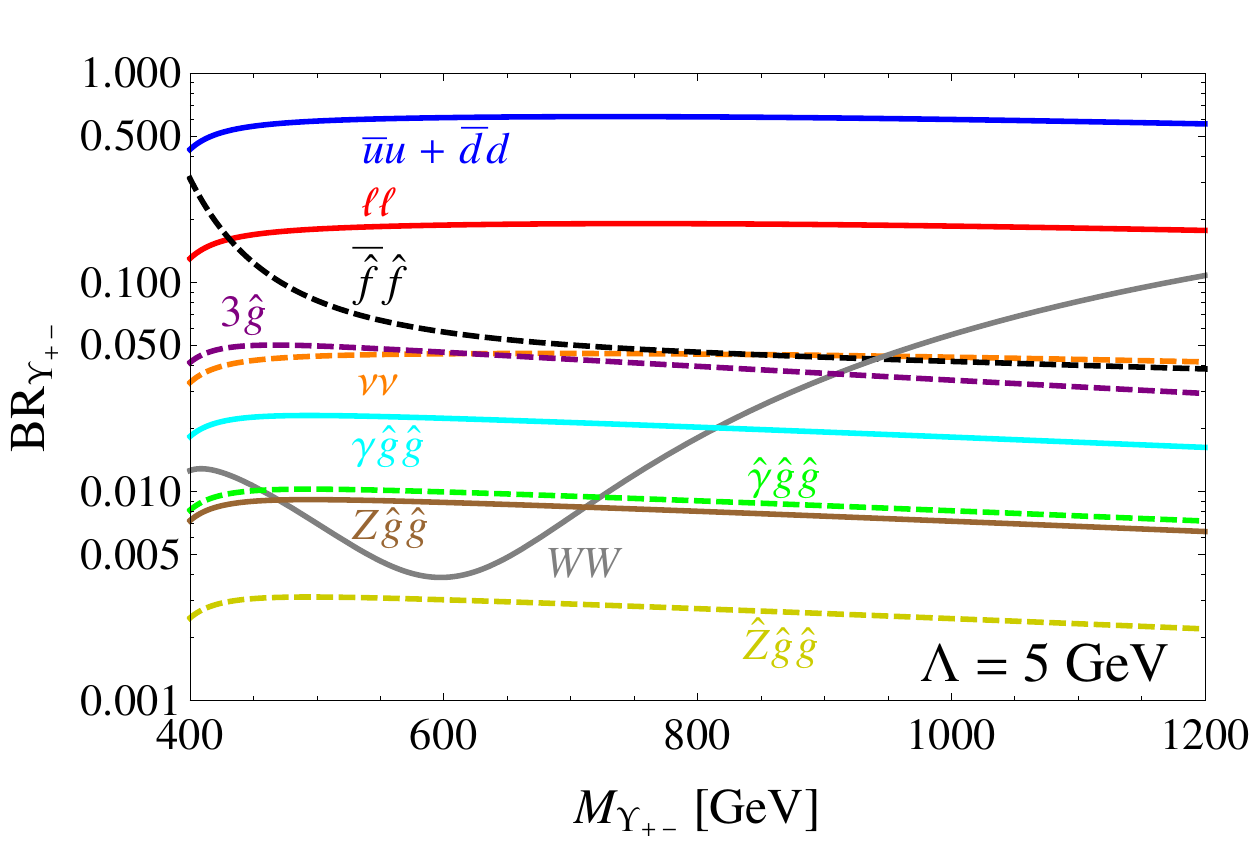}
\caption{Branching ratios of the $\Upsilon_{+-}$ as a function of $M_{\Upsilon_{+-}}\simeq 2\tilde{M}_B$. The twin confinement scale is set to $\Lambda = 5$ GeV. Solid (dashed) curves denote decays into final states containing at least one detectable SM particle (only twin particles or SM neutrinos).}
\label{fig:UpsilonBR}
\end{center}
\end{figure}

\begin{figure}
\begin{center}
\includegraphics[width=8cm]{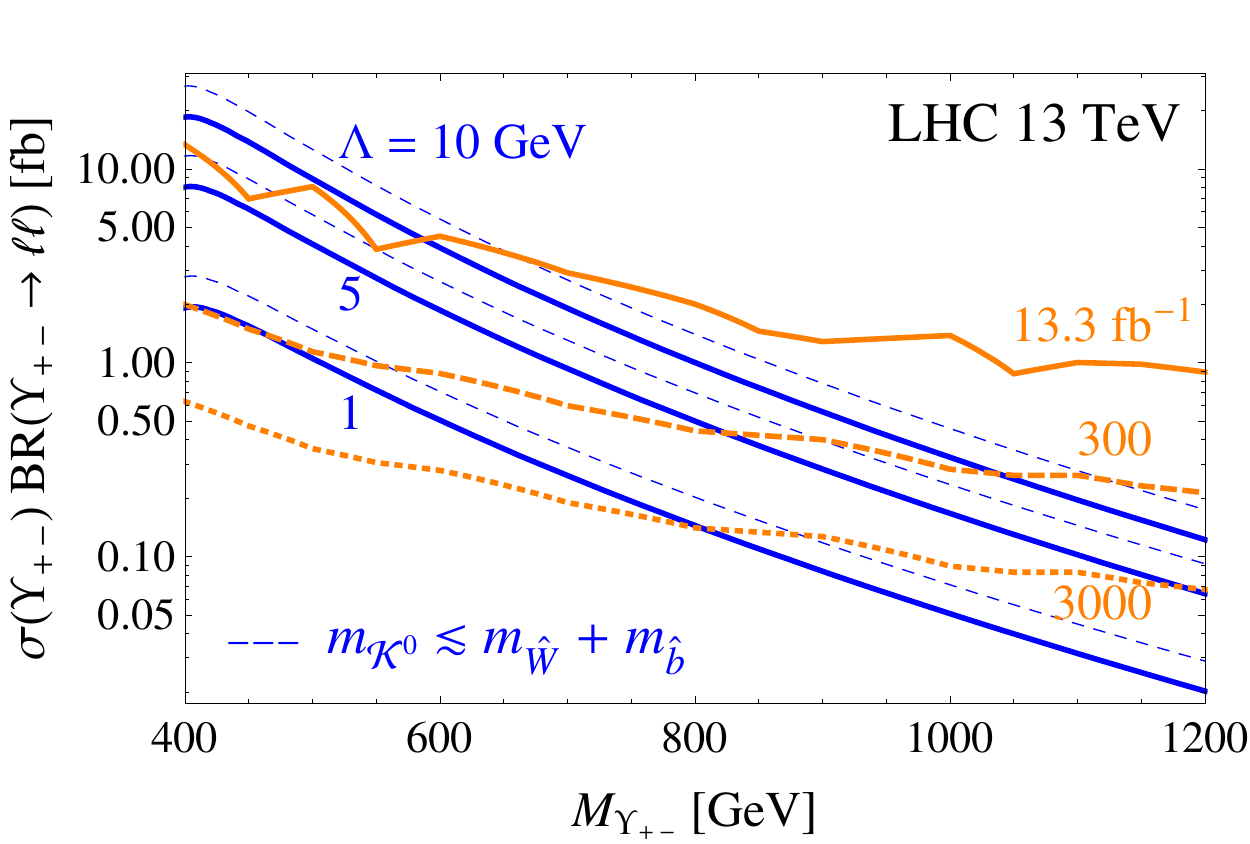}
\includegraphics[width=8cm]{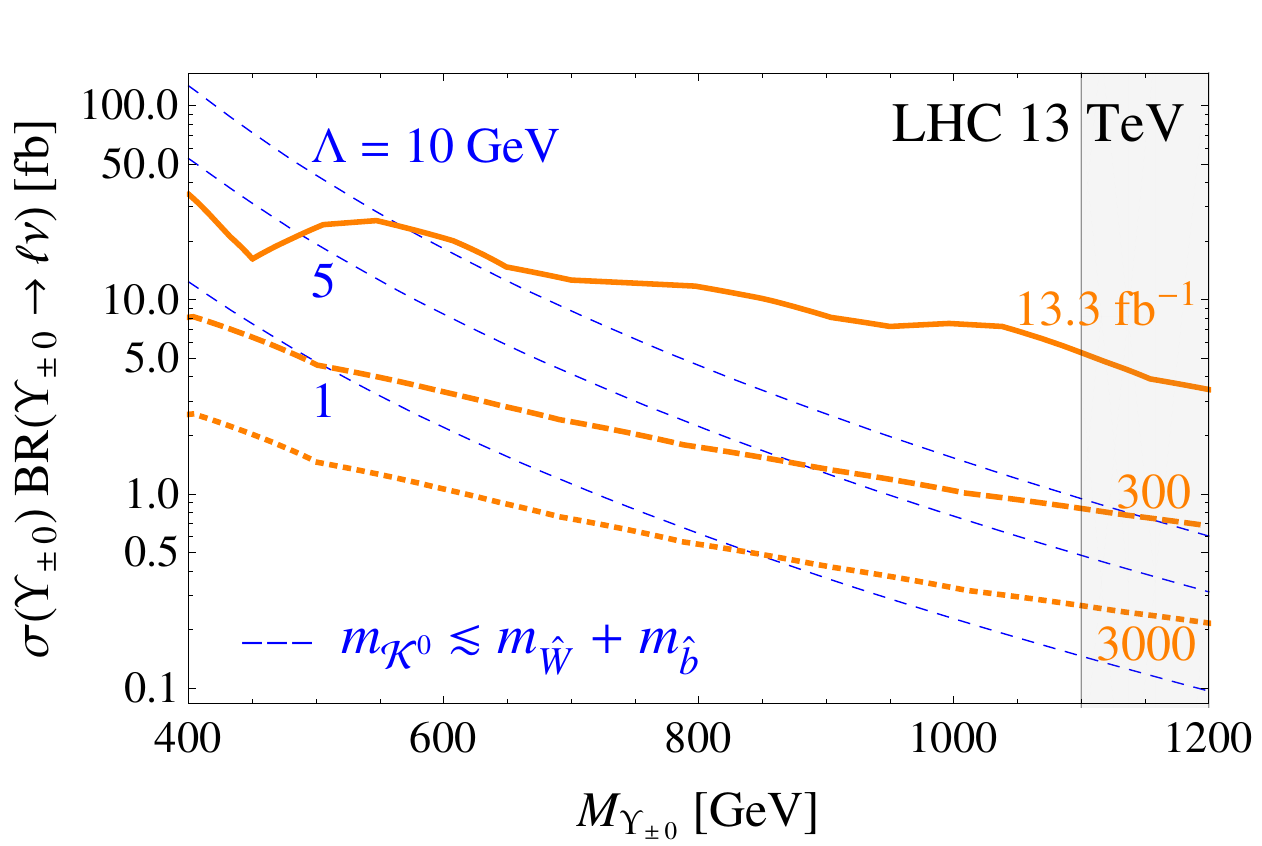}
\caption{{\it Left panel:} in solid blue, the cross section for $\Upsilon_{+-}$ production at the $13$ TeV LHC, multiplied by the branching ratio into one family of SM leptons, as a function of $M_{\Upsilon_{+-}}$. Three different choices of the twin confinement scale $\Lambda$ are considered. The dashed blue lines include also the contribution of $\Upsilon_{00} \to \ell\ell$, which is expected only for $m_{\mathcal{K}^0}\lesssim m_{\hat{W}} + m_{\hat{b}}$; see text for details. In orange, the current constraints from the ATLAS dilepton resonance search (solid), as well as the projections to the end of Run 2 (dashed) and to the High-Luminosity LHC (dotted). {\it Right panel:} similar to the left panel, but for $\Upsilon_{\pm 0}\to \ell\nu$. The $\Upsilon_{\pm 0}$ annihilation signal is expected only if $m_{\mathcal{K}^0} \lesssim m_{\hat{W}} + m_{\hat{b}}$. In the gray-shaded region, this condition requires $m_{\hat{b}}\gtrsim 200$ GeV, which leads to a tuning worse than $10\%$ in the Higgs mass for a cutoff of $5$ TeV. }
\label{fig:UpsilonDileptons}
\end{center}
\end{figure}

What are the main decays of $\Upsilon_{+-}$? As a consequence of twin color conservation, in the perturbative approximation the leading contribution to the twin hadronic width comes from $\Upsilon_{+-} \to 3 \hat{g}$. Thus, as for the SM $J/\psi$ and $\Upsilon$, the strong decays are suppressed, leading to large branching fractions for the decays into fermion-antifermion pairs, mediated by the weak interactions: the $\Upsilon_{+-}$ decays into SM quarks about $60\%$ of the time, see Fig.~\ref{fig:UpsilonBR}.\footnote{There is also a radiative decay $\Upsilon_{+-}\to \eta_{+-} \gamma$, whose width scales as $E_\gamma^3$, where $E_\gamma \simeq M_{\Upsilon_{+-}} - M_{\eta_{+-}} \sim C_{F}^4 \hat{\alpha}_s^4(q_{\rm rms}) m_{\mathcal{K}^-}/3$, leading to a very suppressed branching ratio $\lesssim 10^{-4}$, not shown in Fig.~\ref{fig:UpsilonBR}.}
Experimentally, the most promising decay is into SM dileptons, which provides a very clean final state and benefits from a sizable branching ratio $\mathrm{BR}(\Upsilon_{+-}\to ee+\mu\mu)\simeq 0.12$, with mild dependence on $\tilde{M}_B$. The signal cross section is shown in solid blue in the left panel of Fig.~\ref{fig:UpsilonDileptons}, as a function of $M_{\Upsilon_{+-}} \simeq 2\tilde{M}_B$, for three representative choices of the twin confinement scale: $\Lambda = 5$ GeV is the approximately $Z_2$-symmetric value, while for $\Lambda = 1$ GeV the lightest twin glueball decays far out of the detector, and finally for $\Lambda = 10$ GeV the lightest twin glueball decays promptly \cite{Craig:2015pha}. In the last two cases the displaced decay signatures \cite{Craig:2015pha,Curtin:2015fna} of the twin glueballs are washed out. Then the exotic fermion signals, such as $\Upsilon_{+-}\to \ell\ell$, play an even more important role in probing the TH model at colliders. We compare the signal cross section to the current ATLAS exclusion \cite{ATLAS:2016cyf}, based on $13.3$ fb$^{-1}$ at $13$ TeV, reported as a solid orange curve. Interestingly, a non-trivial constraint, $\tilde{M}_B \gtrsim 300$ GeV, can already be extracted for larger $\Lambda = 10$ GeV. In the same figure we also show the projected constraint after $300$ and $3000$ fb$^{-1}$, obtained by rescaling the current cross section bound $\propto 1/\sqrt{L}$, with $L$ the integrated luminosity. 

The amplitude for the decay of $\Upsilon_{+-}$ into two transverse gauge bosons is velocity-suppressed, and therefore negligible. Thus the only sizable decay width of the spin-$1$ bound state into two gauge bosons is $\Upsilon_{+-} \to WW$ (depicted in the middle diagram of Fig.~\ref{fig:Upsilon decay}), mediated by the coupling of $\mathcal{K}^-$ to the longitudinal $W$, which originates from the top Yukawa
\begin{equation} \label{longW coupling}
\mathcal{L}_t \;\;\; \ni \;\;\; i y_t\, \pi^+ \bar{u}_{3R}^B \mathcal{K}^-_L + \mathrm{h.c.},
\end{equation}
where $\mathcal{L}_t$ was defined in Eq.~\eqref{eq:yukawa}.

The total width of the $\Upsilon_{+-}$ is of $O(1$-$10)$ MeV, which on the one hand is much smaller than the LHC experimental resolution, and on the other hand satisfies $\Gamma_{\Upsilon_{+-}} \gg \Gamma_{\mathcal{K}^-}$ across all parameter space considered here, guaranteeing that the bound state annihilation occurs before the individual constituents can decay.      

The pseudoscalar $\eta_{+-}$ is produced in photon fusion, with cross section
\begin{align} \label{gammagamma}
\sigma_{\gamma\gamma}(pp \to \eta_{+-}) \,=&\, 8\pi^2 \frac{\Gamma^{\eta_{+-}}_{\gamma\gamma}}{2 m_{\mathcal{K}^-}}\frac{L_{\gamma\gamma}(\frac{4m_{\mathcal{K}^-}^2}{s})}{s}
\end{align}
where 
\begin{equation} \label{diphoton width}
\Gamma^{\eta_{+-}}_{\gamma\gamma} = N_c 4\pi \alpha^2 Q_{\mathcal{K}^-}^4\frac{\left|\psi(0)\right|^2}{m_{\mathcal{K}^-}^2}\,,
\end{equation}
is the width for $\eta_{+-}\to \gamma\gamma$, and $L_{\gamma\gamma}(\tau) = \int_\tau^1 \frac{dx}{x} \gamma(x) \gamma(\tau/x)$ is the $\gamma\gamma$ luminosity \cite{Manohar:2016nzj}. Numerically, the production of $\eta_{+-}$ is roughly two orders of magnitude smaller than that of $\Upsilon_{+-}$: for the benchmark $\tilde{M}_B = 350$ GeV, $\sigma_{\gamma\gamma}(pp\to \eta_{+-})$ varies from $0.033$ to $0.26$ fb for $\Lambda\in [1,10]$ GeV. In addition, the $\eta_{+-}$ width is dominated by decays into twin hadrons. In the perturbative approximation, these can be parameterized by the $\eta_{+-}\to \hat{g}\hat{g}$ process, which has a branching ratio of $0.8\,$-$\,0.9$ across the parameter space. This suppresses the branching ratios into SM final states, among which $\gamma\gamma$ and $WW$ would be most promising experimentally, to the few percent level. Thus detection of the $\eta_{+-}$ appears challenging even with high luminosity. The total width of $\eta_{+-}$ is a factor $\sim 5\,$-$\,10$ larger than $\Gamma_{\Upsilon_{+-}}$ due to the unsuppressed decay into twin hadrons.    

We conclude this subsection with a comment on the validity of the Coulomb approximation for the bound states, which applies if
\begin{equation} \label{CoulombBreakdown}
\frac{a_0}{\Lambda^{-1}}\ll 1\,,
\end{equation}
where $a_0$ is the Bohr radius, defined below Eq.~\eqref{wavefunctionCoulomb}. In the entire parameter space shown in Fig.~\ref{fig:UpsilonDileptons} we have $a_0 / \Lambda^{-1} \lesssim 0.2$, confirming the applicability of the Coulomb treatment. For even larger $\Lambda$ the effects of confinement become important, and the approximation will not be valid.

\subsection{Bound states containing $\mathcal{K}^0$}\label{subs Z2break:2}
Next, we turn to the possible observation of bound states containing $\mathcal{K}^0$, in particular the spin-1 $(\bar{\mathcal{K}}^0 \mathcal{K}^0)$ state, labeled by $\Upsilon_{00}$, and the electrically charged $(\bar{\mathcal{K}}^0\mathcal{K}^-)$ vector, $\Upsilon_{-0}$. If $\mathcal{K}^0$ decays into on-shell $\hat{W}\hat{b}$, its lifetime is in the range $1\;\mathrm{MeV} \lesssim \Gamma_{\mathcal{K}^0} \lesssim 1\;\mathrm{GeV}$. Since (barring $O(1)$ factors) $\Gamma_{\Upsilon_{00}} \sim \Gamma_{\Upsilon_{-0}} \sim \Gamma_{\Upsilon_{+-}} = O(1$-$10)$ MeV, this implies that the constituent $\mathcal{K}^0$'s likely decay before the bound states they form can annihilate. Conversely, for $m_{\mathcal{K}^0} \lesssim m_{\hat{W}} + m_{\hat{b}}$ we find $\Gamma_{\mathcal{K}^0} \lesssim 1$ MeV, which guarantees signals from $\Upsilon_{00}$ and $\Upsilon_{\pm 0}$. In the remainder of this subsection we discuss the corresponding phenomenology.

The $\Upsilon_{\pm 0}$ is produced in the charged DY channel, and decays primarily into SM fermion pairs. Subleading decays are into $W\hat{g}\hat{g}$ and into $WV_0$, where $V_0$ is a neutral electroweak gauge boson. The $\Upsilon_{\pm 0} \to WV_0$ widths arise from the interaction in Eq.~\eqref{longW coupling}, after $\mathcal{K}^0$-$\,\hat{t}$ mixing. The mixing also implies that the mass of $\Upsilon_{\pm 0}$ is expected to be slightly smaller than that of $\Upsilon_{+-}$, since $m_{\mathcal{K}^0} < m_{\mathcal{K}^-}$. The most promising signal is $\Upsilon_{\pm 0} \to \ell \nu$, whose cross section at the $13$ TeV LHC is shown in the right panel of  Fig.~\ref{fig:UpsilonDileptons}, together with the constraint from the current ATLAS $\ell\nu$ resonance search \cite{ATLAS:2016ecs} and its projections to the end of Run 2 and to the HL-LHC. For the sake of simplicity, in this figure the $\mathcal{K}^0$-$\,\hat{t}$ mixing was neglected, leading to the simplifications $M_{\Upsilon_{\pm 0}} \simeq 2\tilde{M}_B$ and $\Gamma(\Upsilon_{\pm 0} \to WV_0) = 0$. On the other hand, if the decay $\mathcal{K}^0 \to \hat{W}\hat{b}$ is fast, the electroweak production of a $\bar{\mathcal{K}}^0\mathcal{K}^-$ pair is followed by the formation of a $(\bar{\hat{b}}\mathcal{K}^-)$ bound state instead. Provided its angular momentum is not too large, this bound state could annihilate on a time scale shorter than the $\mathcal{K}^-$ lifetime. Electric charge conservation forces the annihilation channel to be $W^- \hat{W}^+$, thus giving rise to the $W$+MET signature.

The $\Upsilon_{00}$ phenomenology is qualitatively similar to that of the $\Upsilon_{+-}$. One important quantitative difference is that the mass of $\Upsilon_{00}$ is $(M_{\Upsilon_{+-}}-M_{\Upsilon_{00}})/M_{\Upsilon_{+-}} \simeq 1 - m_{\mathcal{K}^0}/m_{\mathcal{K}^-}$ smaller, which increases from $3\%$ at $M_{\Upsilon_{+-}} = 400$ GeV to $8\%$ at $M_{\Upsilon_{+-}} = 1200$ GeV for $f=1$~TeV. For comparison, the experimental resolution at dilepton invariant mass of $1$ TeV is $1\%$ in the $ee$ channel and $6\%$ in the $\mu\mu$ channel \cite{ATLAS:2016cyf}, suggesting that the two-peak structure formed by the $\Upsilon_{+-}$ and $\Upsilon_{00}$ would likely be resolved in the dielectron channel, while a single broader peak would be observed in dimuons. For simplicity, here we ignore the $\mathcal{K}^0$-$\,\hat{t}$ mixing that generates the small mass splitting between the resonances, and simply show in the left panel of Fig.~\ref{fig:UpsilonDileptons} (dashed blue) the sum of the $\Upsilon_{+-}$ and $\Upsilon_{00}$ dilepton signals as a function of $M_{\Upsilon_{+-}}\simeq M_{\Upsilon_{00}}$. The $\Upsilon_{00}$ contributes about half compared to the $\Upsilon_{+-}$, due to a smaller branching ratio into dileptons. A comparison of the two panels of Fig.~\ref{fig:UpsilonDileptons} shows that for $m_{\mathcal{K}^0} \lesssim m_{\hat{W}} + m_{\hat{b}}$, the reach in the $\Upsilon_{+-,\,00}\to \ell\ell$ and $\Upsilon_{\pm 0} \to \ell\nu$ final states is similar.

We conclude with a comment on the pseudoscalar states $\eta_{00}$ and $\eta_{\pm 0}$, whose production cross sections at the LHC are tiny. In fact, since $\mathcal{K}^0$ has zero electric charge, $\eta_{00}$ is not produced in photon fusion, while the production of $\eta_{\pm0}$ in $W\gamma$ fusion is very suppressed by the small $W$ PDF.

\subsection{The role of a light exotic vector}\label{subs Z2break:3}
So far we have neglected the effects of the $SU(6)$ off-diagonal vector $\mathcal{X}$ on the physics of the bound states composed of exotic fermions. However, since the coupling $g_{\mathcal{X}}$ is expected to be naturally rather large, a relatively light $\mathcal{X}$ can have significant effects on the bound state phenomenology, which will be discussed here.

\begin{figure}
\begin{center}
\includegraphics[width=8cm]{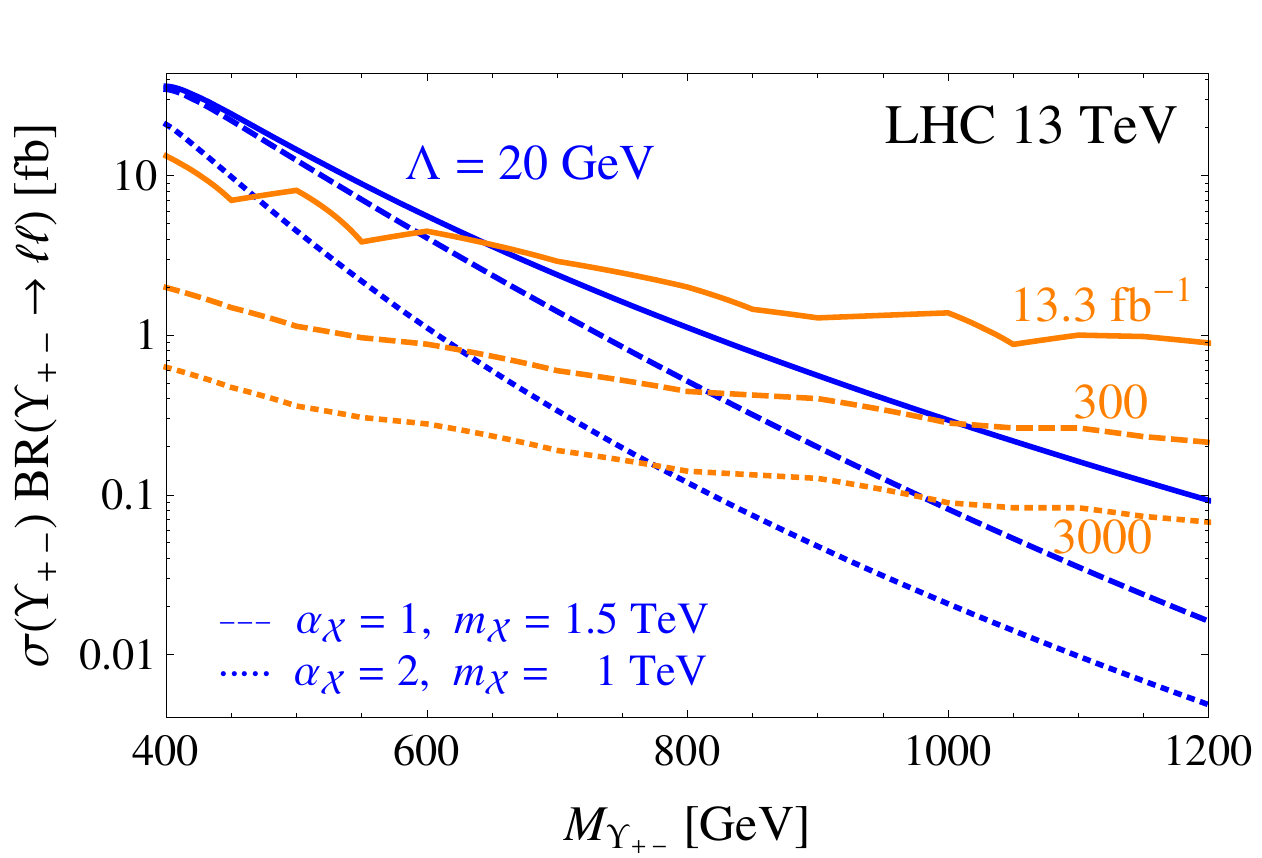}
\includegraphics[width=8cm]{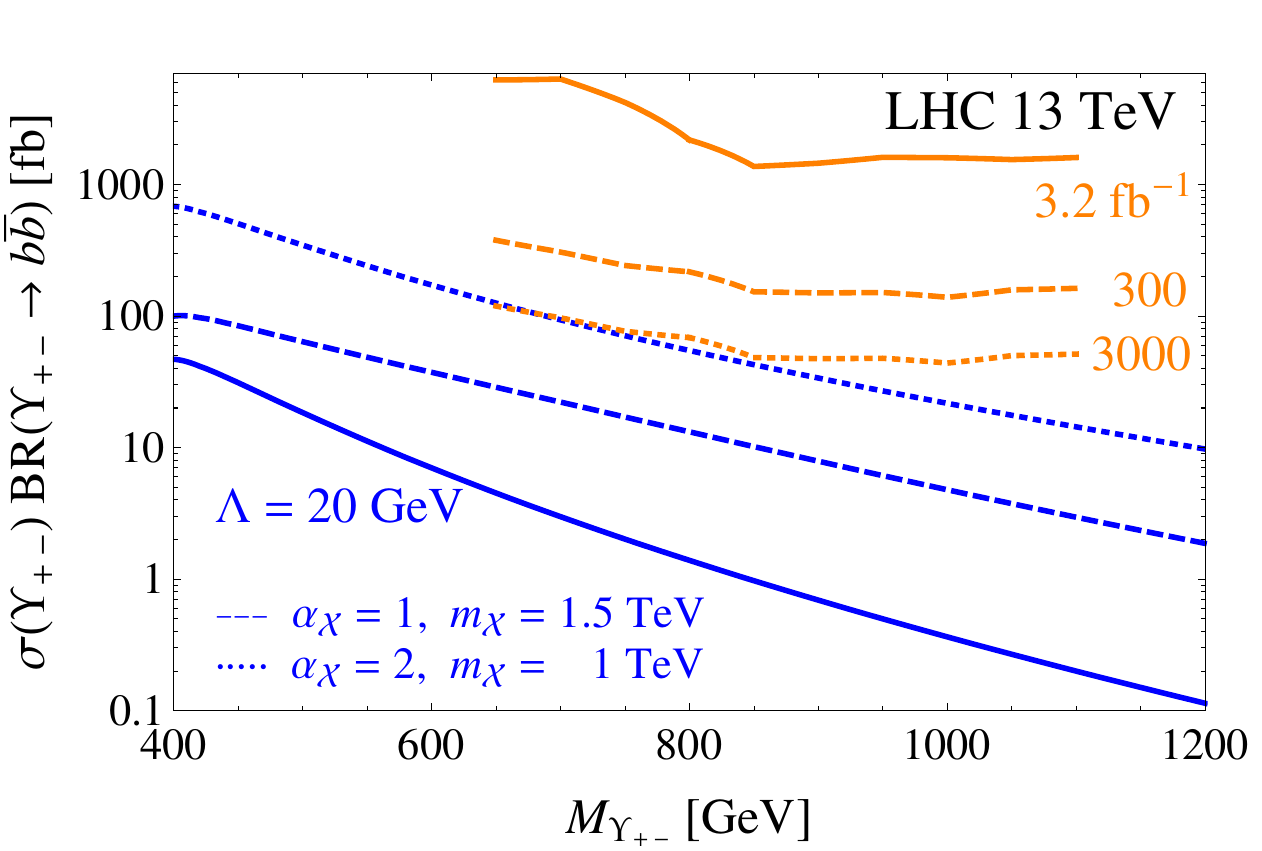}
\caption{In the left (right) panel, the solid blue curve indicates the $\ell\ell$ $(b\bar{b})$ signal from the $\Upsilon_{+-}$, for twin confinement scale $\Lambda = 20$ GeV. The dashed and dotted blue curves show the cross section in presence of a light exotic vector. The orange curves show the current and projected experimental constraints, similarly to Fig.~\ref{fig:UpsilonDileptons}.}
\label{fig:UpsilonLambda20}
\end{center}
\end{figure}

Through the coupling in Eq.~\eqref{exovector coupl}, the $t$-channel exchange of the exotic vector mediates the decay of $\Upsilon_{+-}$ into $b\bar{b}$, which is depicted in the last diagram in Fig.~\ref{fig:Upsilon decay}. The width is
\vspace{0.5cm}
\begin{equation} \label{vector to bb}
\Gamma_{b\bar{b}}^{\Upsilon_{+-}} = N_c^2 \frac{\pi \alpha_{\mathcal{X}}^2}{6} \frac{\left|\psi(0)\right|^2}{m_{\mathcal{K}^-}^2}\frac{m_{\mathcal{K}^-}^4}{(m_{\mathcal{X}}^2 + m_{\mathcal{K}^-}^2)^2}\left(1 + \frac{m_{\mathcal{K}^-}^2}{2m_{\mathcal{X}}^2}\right)^2\,,
\vspace{0.5cm}
\end{equation}
\enlargethispage{-50pt}
where the interference with the $s$-channel $\gamma/Z$ exchange was neglected. Thus for sufficiently light $\mathcal{X}$ the $\Upsilon_{+-}$ acquires a large branching fraction into $b\bar{b}$, which in particular dilutes the dilepton rate. At the same time, the exotic vector also mediates $\Upsilon_{+-}$ production in $b\bar{b}$ annihilation, with cross section given by a formula similar to Eq.~\eqref{drellyan}. The total effect on the dilepton signal is shown in the left panel of Fig.~\ref{fig:UpsilonLambda20}, where $\Lambda = 20$ GeV was assumed.\footnote{For $\Lambda = 20$ GeV the condition in Eq.~\eqref{CoulombBreakdown} is not satisfied, hence the bound states cannot be described within the Coulomb approximation. In this case we replace Eq.~\eqref{wavefunctionCoulomb} with the ansatz $\left|\psi(0)\right|^2/m_{\mathcal{K}^-}^3 = \Lambda^2/m_{\mathcal{K}^-}^2$, which is motivated by the physical picture of a string with linear potential, and holds to reasonable accuracy for the lowest-lying SM bottomonia. For example, from the measured value of the electronic width of the $\Upsilon$ \cite{PDG} one extracts $\left|\psi(0)\right|^2/m_b^3 \simeq 3.7\,(2.8)\times 10^{-3}$, where the number in parentheses was obtained using our ansatz with $\Lambda_{\rm QCD} = 250$ MeV.} 
For this larger confinement scale, which is at the upper edge of the plausible range in the Fraternal TH \cite{Craig:2015pha}, the increased $\Upsilon_{+-}$ production cross section implies that masses $M_{\Upsilon_{+-}} \lesssim 650$ GeV are ruled out already at the time of writing, and the HL-LHC will be able to probe up to $M_{\Upsilon_{+-}} \sim 1.2$ TeV. However, as shown by Fig.~\ref{fig:UpsilonLambda20}, an exotic vector with mass $O(\mathrm{TeV})$ and coupling $g_{\mathcal{X}}\sim 3\,$-$\,5$ can suppress the dilepton rate significantly. This is accompanied by a large increase of the $\Upsilon_{+-} \to b\bar{b}$ signal, shown in the right panel of Fig.~\ref{fig:UpsilonLambda20} together with the current \cite{ATLAS:2016fol} and projected experimental constraint.\footnote{To compare our theoretical prediction with the experimental constraint reported in Ref.~\cite{ATLAS:2016fol} we used the following expressions for the $b$-tagging efficiency and kinematic acceptance: $\epsilon = 0.22 - 0.36(m_{b\bar{b}}/\mathrm{TeV}-0.65)$ and $\mathcal{A} = 0.22 + 0.88(m_{b\bar{b}}/\mathrm{TeV}-0.65)$ for $m_{b\bar{b}}<0.8$ TeV, $\mathcal{A} = 0.35$ for $m_{b\bar{b}}> 0.8$ TeV, valid in the range $m_{b\bar{b}}\in [0.65,1.1]$ TeV. We thank A.~Coccaro for useful clarifications about Ref.~\cite{ATLAS:2016fol}.} Interestingly, even for a very light and strongly coupled $\mathcal{X}$ the $\Upsilon_{+-}$ resonance in the $b\bar{b}$ final state would remain hidden after the HL-LHC. Notice also that, since $\mathcal{X}$ only couples to the left-handed $\tilde{q}_3^B$, the exotic vector exchange does not mediate the decay of the pseudoscalar $\eta_{+-}$ into bottom quarks.

\enlargethispage{-100pt}

A relatively light $\mathcal{X}$ also provides, through its decays, an additional mechanism for production of exotic fermion bound states. In particular, the process $pp\to (\mathcal{X}\to b\bar{\mathcal{K}}^+)(\mathcal{X}^\ast \to \bar{b}\mathcal{K}^-)$, where the $\mathcal{X}\mathcal{X}^\ast$ production is via QCD (see Eq.~\eqref{gg to XX}), is followed by formation of a twin QCD string between the $\mathcal{K}^-$ and $\bar{\mathcal{K}}^+$. Based on a simple counting of degrees of freedom, we can estimate that the resulting bound state will be a $\Upsilon_{+-}$ with probability $\sim 3/4$, or a $\eta_{+-}$ with probability $\sim 1/4$. Then the dominant final state is expected to be $4b$, stemming from $b\bar{b}(\Upsilon_{+-}\to b\bar{b})$. Notice that the total width of the exotic vector is $\Gamma_{\mathcal{X}} = \alpha_{\mathcal{X}} m_{\mathcal{X}}/4$ (where we have assumed that $\tilde{M}_A$ is too large for $\mathcal{X}$ to decay into the exotic quarks $\tilde{q}_3^A$), which for $\alpha_{\mathcal{X}} \sim O(1)$ implies a broad resonance. 

Although phenomenologically interesting, a light $\mathcal{X}$ is however potentially problematic for two reasons. First, it should be accompanied by a light KK gluon, which is subject to strong experimental constraints. Second, the $Z_2$ breaking $\tilde{M}_A \neq \tilde{M}_B$ leads to a $2$-loop quadratically divergent correction to the Higgs mass that is negligible for large $m_{\mathcal{X}}$, but can become important if the exotic vector is light, thus increasing the fine-tuning. These aspects are discussed in the next two subsections.

\subsection{KK gluon}\label{subs Z2break:4}
In the minimal version of the two-site model, from the kinetic term of the link field $\Sigma_6$ \footnote{We denote the $SU(6)$ link field by $\Sigma_6$, to distinguish it from the $SU(4)$ one that we labeled $\Sigma$.} one finds that the mass of the KK gluon $\mathcal{G}$ is related to the mass of $\mathcal{X}$ by $m_\mathcal{G}= m_\mathcal{X}/\cos \theta_A$, where $\sin \theta_A = g_s/g_{\mathcal{X}}$ (see App.~\ref{SU4exo}). Because $g_\mathcal{X} \gg g_s$, this would give $\mathcal{G}$ a mass similar to, but slightly higher than $m_\mathcal{X}$. However, this relation can be violated by higher dimensional operators. For example, $|\mathrm{Tr} (\Sigma_6^\dagger D_\mu \Sigma_6\, \Omega)|^2$ with a spurion $\Omega = {\rm diag}( \mathbf{1}_3, \mathbf{0}_3)$ inserted on the $SU(3)^2$ site makes $m_\mathcal{G}^2$ twice as big as $m_\mathcal{X}^2$ (if one ignores $g_s$). In summary, $m_\mathcal{G}$ is expected to be of the same order as $m_\mathcal{X}$, although with some freedom. Since the KK gluon, being a SM color octet, has a very large production cross section at hadron colliders (see Fig.~\ref{fig:RRxsec}), it is then important to take it into account in the discussion of a light $SU(6)$ exotic vector.  

\begin{figure}
\begin{center}
\includegraphics[width=8cm]{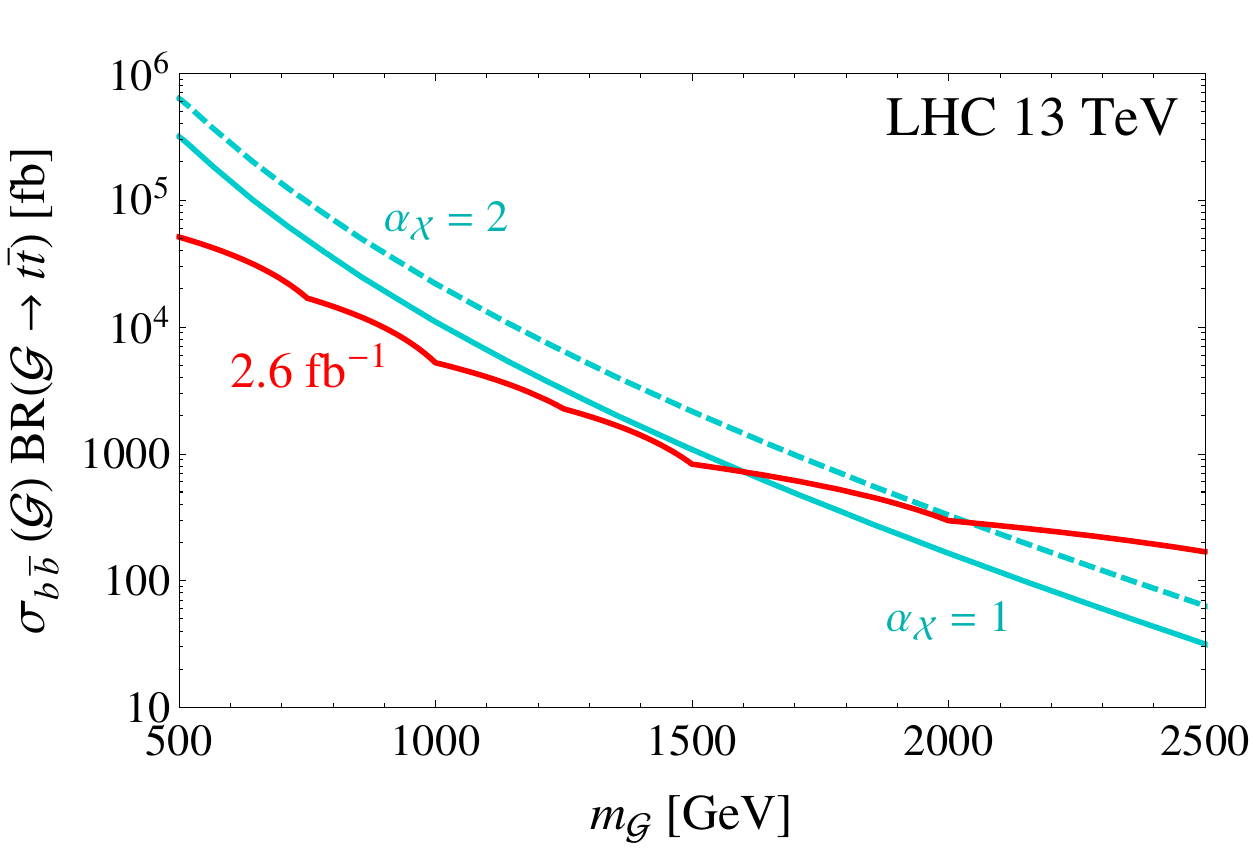}
\caption{In light blue, the cross section for single KK gluon production in $b\bar{b}$ annihilation, multiplied by the branching ratio into $t\bar{t}$, for two representative values of $\alpha_{\mathcal{X}}$. The MSTW08LO PDFs were used, evaluated at $\mu_{\rm fact} = m_{\mathcal{G}}\,$. In red, the CMS bound from the first $13$ TeV data.}
\label{fig:KKgluontt}
\end{center}
\end{figure}

The partonic cross section for $gg\to \mathcal{G}\mathcal{G}$, which mediates KK gluon pair production, can be found for example in Ref.~\cite{Dobrescu:2007yp}, and has a structure similar to Eq.~\eqref{gg to XX}. The couplings of $\mathcal{G}$ to fermions read approximately
\begin{equation} \label{KK gluon fermions}
g_{\mathcal{X}} \mathcal{G}_\mu^a ( \bar{q}_{3L}^{A\,i} \gamma^\mu t^a_{ij} q_{3L}^{A\,j} + \bar{u}_{3R}^{A\,i} \gamma^\mu t^a_{ij}u_{3R}^{A\,j} ),
\end{equation} 
whereas the coupling to the $b_R$ is strongly suppressed by $g_s/g_{\mathcal{X}}$. Therefore the KK gluon decays into $t\bar{t}$ and $b\bar{b}$ with approximate branching ratios $2/3$ and $1/3$, respectively. Since the total width is $\Gamma_{\mathcal{G}} \simeq \alpha_{\mathcal{X}} m_{\mathcal{G}}/4$, the KK gluon, like the $\mathcal{X}$, is expected to be a broad resonance. The leading constraint on $\mathcal{G}$ pair production comes from the $4t$ final state, where the $51$ fb upper limit on the four-top cross section \cite{ATLAS:2016btu} translates into $m_{\mathcal{G}} \gtrsim 1.3$ TeV.\footnote{Reference~\cite{ATLAS:2016btu} quotes bounds on the four-top cross section assuming several production mechanisms. The $51$ fb limit is for the case of a $4t$ contact interaction, which we believe to be the best available approximation, because $\mathcal{G}$ is a broad resonance.} However, the KK gluon can also be singly produced in $b\bar{b}$ annihilation, with cross section
\begin{equation}
\sigma_{b\bar{b}}(pp\to \mathcal{G}) = \frac{8\pi^2  \alpha_{\mathcal{X}}}{9} \frac{L_{b\bar{b}}\left(\frac{m_{\mathcal{G}}^2}{s}\right)}{s}\, .
\end{equation}
The large coupling $g_{\mathcal{X}}$ can compensate the suppressed $b\bar{b}$ PDF, leading to competitive constraints from searches for $t\bar{t}$ resonances. From the CMS analysis in Ref.~\cite{CMS:2016zte} we obtain $m_{\mathcal{G}}\gtrsim 1.6\,(2.0)$ TeV for $\alpha_{\mathcal{X}} = 1\,(2)$, see Fig.~\ref{fig:KKgluontt}.

The couplings of $\mathcal{G}$ to the third-generation SM fermions, Eq.~\eqref{KK gluon fermions}, could also lead to constraints from flavor observables. We focus on the down-quark sector, which gives the strongest bounds. Integrating out $\mathcal{G}$ and applying Fierz transformations, we obtain the effective Lagrangian
\begin{equation}
\mathcal{L}_{\rm eff} = - \frac{g_{\mathcal{X}}^2}{6m_{\mathcal G}^2} \bar{b}_L^i \gamma^\mu b_L^i \bar{b}_L^{j}\gamma_\mu b_L^j \,.
\end{equation}
We observe that since the coupling of $\mathcal{G}$ to the $b_R$ is very suppressed, only LL operators are generated with relevant size, whereas experimentally, the strongest bounds are on the coefficients of LR operators. Applying to the down-type quarks the rotation that diagonalizes the mass matrix and selecting the $\Delta S = 2$ operator that contributes to Kaon mixing, we find (see for example Ref.~\cite{Csaki:2008zd})
\begin{equation} \label{eff operator}
\mathcal{H}_{\rm eff} = \frac{g_{\mathcal X}^2}{6m_{\mathcal G}^2} |V_{td} V_{ts}|^2 \bar{d}_L \gamma^\mu s_L \bar{d}_L\gamma_\mu s_L\,,
\end{equation}
where $V_{ij}$ are elements of the Cabibbo-Kobayashi-Maskawa matrix, and we assumed the coefficient of the operator to be real. Defining the coefficient of $\bar{d}_L \gamma^\mu s_L \bar{d}_L\gamma_\mu s_L$ as $1/\Lambda_{\rm F}^2$, the experimental bound reads \cite{UTfit} $\,\Lambda_{\rm F} > 1.0\times 10^3\;\mathrm{TeV}$. This can be translated into $m_{\mathcal G}/g_{\mathcal X} > 0.14$ TeV, which is weaker than the LHC constraints for a typical coupling $g_{\mathcal X} \sim 3\,$-$\,5$. Notice, however, that if the operator in Eq.~\eqref{eff operator} were generated with an unsuppressed phase, the bound on the imaginary ($CP$-violating) part would read $\Lambda_{\rm F} > 2.6\times 10^4\;\mathrm{TeV}$ \cite{Ligeti:2016qpi}, leading to $m_{\mathcal{G}}/g_{\mathcal X} > 3.7$ TeV. Thus, to envisage a KK gluon lighter than $\sim 10$ TeV we need to postulate that its couplings respect $CP$ invariance \cite{Csaki:2008zd}.

\subsection{Implications for naturalness}\label{subs Z2break:5}
Since the $Z_2$ symmetry is crucial in protecting the Higgs potential from large radiative corrections, a $Z_2$-breaking spectrum for the exotic fermions necessarily has implications on naturalness. The contribution to the potential from the top sector of Eqs.~(\ref{eq:yukawa}),~(\ref{eq:mass}) is finite for $\tilde{M}_A = \tilde{M}_B$ \cite{Chacko:2005pe}, but if this relation is violated then the logarithmic divergences do not cancel exactly, leading to an additional contribution
\begin{align} \label{1loop Z2break}
V_{\rm top}^{\slashed{Z_2}}& - V_{\rm top}^{Z_2} = \nonumber \\ \,\, -&\,\frac{3y_t^2}{16\pi^2} h^2\,\,\tfrac{\tilde{M}_A^4 \left(\tilde{M}_B^2- y_t^2 f^2\right) \text{log}\frac{\Lambda_{\rm UV}^2}{\tilde{M}_A^2}\,-\,(A \leftrightarrow B)\,+\,\left(\tilde{M}_A^2-\tilde{M}_B^2\right) \left[\left(\tilde{M}_A^2- y_t^2 f^2\right) \left(\tilde{M}_B^2 - y_t^2 f^2\right)+y_t^4 f^4  \text{log}\frac{\Lambda_{\rm UV}^2}{y_t^2 f^2 }\right]}{\left(\tilde{M}_A^2 - y_t^2 f^2\right) \left(\tilde{M}_B^2 -  y_t^2 f^2\right)} + O(h^4),
\end{align}
where $\Lambda_{\rm UV}$ is the cutoff. Unless the splitting between $\tilde{M}_A$ and $\tilde{M}_B$ is very large, this extra term is of moderate size and does not require significantly more tuning. Notice also that in the region of parameters studied here, $\tilde{M}_A > \tilde{M}_B$, the sign of the extra mass term is negative and thus does not help to achieve $\left\langle h \right\rangle \ll f$, which is needed for a realistic model. An additional, small source of $Z_2$ breaking is therefore still required. In passing we note that in the opposite regime $\tilde{M}_A < \tilde{M}_B$, the $Z_2$ breaking from the exotic fermion masses may be sufficient for realistic EWSB. 

Closer inspection reveals that the $Z_2$-breaking $\tilde{M}_A \neq \tilde{M}_B$ also gives rise to a $2$-loop quadratic divergence in the Higgs mass. This can be understood by noticing that due to the couplings in Eq.~\eqref{exovector coupl}, the wavefunction of $u_{3L}^A$ is renormalized by a loop of $\mathcal{X}$ and $\tilde{u}_{3L}^B$, $i\bar{u}_{3L}^A \slashed{\partial}u_{3L}^A \to i(1 + \delta Z_L)\bar{u}_{3L}^A \slashed{\partial}u_{3L}^A$, whereas the wavefunction of $u_{3L}^B$ is renormalized by a loop of $\mathcal{X}$ and $\tilde{u}_{3L}^A$, with renormalization constant $\delta\hat{Z}_L$. For $\tilde{M}_A \neq \tilde{M}_B$ we have $\delta Z_L \neq \delta\hat{Z}_L$, leading to a $2$-loop quadratic contribution to the physical Higgs mass,
\begin{align} 
\delta m_h^2 \,=&\, \frac{3y_t^2}{4\pi^2}\Lambda_{\rm UV}^2\, (\delta \hat{Z}_L - \delta Z_L) \nonumber \\ \,\sim &\, \frac{3y_t^2}{4\pi^2}\,\Lambda_{\rm UV}^2\, \frac{3N_c \alpha_{\mathcal X} }{16\pi}\,\frac{\tilde{M}_B^2 - \tilde{M}_A^2}{m_{\mathcal X}^2}\,\log \frac{\Lambda_{\rm UV}^2}{m_{\mathcal X}^2}\,. \label{2loop Z2break}
\end{align}
For light $\mathcal{X}$, this correction is important and an increased fine-tuning is required to obtain $m_h = 125$ GeV: for example, taking $m_{\mathcal{X}} = 1.5$ TeV, $\alpha_{\mathcal X} = 1$, $\tilde{M}_{B} = \tilde{M}_A/2 = 500\;\mathrm{GeV}$ and a cutoff $\Lambda_{\rm UV} = 5$ TeV, we find a tuning of $\sim 10\%$ (we take $y_t^2 \sim 1/2$ at scale $\Lambda_{\rm UV}$).\footnote{Notice that to obtain the last equality in Eq.~\eqref{2loop Z2break} we have assumed $m_{t, \hat{t}} \ll \tilde{M}_{A,B} \ll m_{\mathcal X}$, since in this limit the result is most transparent. In our actual parameter space we have $\tilde{M}_B < m_{\hat{t}}$, but Eq.~\eqref{2loop Z2break} still provides a reasonably accurate estimate of the fine-tuning.} This shows that large effects of the exotic vector on the physics of the exotic fermion bound states, such as those illustrated by Fig.~\ref{fig:UpsilonLambda20} in \ref{subs Z2break:3}, would be associated with a stronger fine-tuning. It is important to stress, however, that the bound state signals discussed in \ref{subs Z2break:1} and \ref{subs Z2break:2} do not need to carry a naturalness price tag, because the $\mathcal{X}$ may simply be heavy, strongly suppressing the correction in Eq.~\eqref{2loop Z2break}.

An independent $2$-loop tuning arises if the twin confinement scale $\Lambda$ differs from the $Z_2$-symmetric value $\simeq 5$ GeV \cite{Craig:2015pha}. This requires a splitting of the standard and twin QCD couplings at high scale, $\hat{g}_s(\Lambda_{\rm UV}) \neq g_s(\Lambda_{\rm UV})$. The associated $2$-loop contribution to the Higgs mass can be roughly estimated as \cite{Craig:2015pha}
\begin{equation} \label{2loopTuning}
\delta m_h^2 \sim {\frac{3y_t^2}{8\pi^4}(g_s^2 - \hat{g}_s^2)\,\Lambda^2_{\rm UV}}\,.
\end{equation}
For reference, if twin QCD is not gauged ($\hat{g}_s = 0$) this estimate gives $\sim 25\%$ tuning, where we used $\alpha_s \sim 0.07$ at high scale.

\section{Conclusions}\label{conclusions}
In this paper we studied the collider phenomenology of the exotic states which appear in the non-supersymmetric UV completions of the TH model. These exotic states carry charges under both the SM and twin gauge groups, so they constitute a window to the twin sector. They may also provide the first experimental hint of the TH, given that the low energy effective theory is quite elusive and may even escape future LHC detection for some range of underlying parameters. The exotic quarks $\tilde{q}_3^A$ that carry the SM color and twin electroweak charges were the subject of a previous study~\cite{Cheng:2015buv}. Here we extended the analysis to several other states, including the exotic fermions $\tilde{q}_3^B$, which are charged under the SM electroweak and twin color groups, as well as the exotic vectors $\mathcal{W}$ and $\mathcal{X}$, which transform in the bifundamental of $SU(2)_A \times SU(2)_B$ and $SU(3)_A \times SU(3)_B$, respectively. As in Ref.~\cite{Cheng:2015buv}, due to naturalness considerations the low energy theory was assumed to be approximately described by the Fraternal TH model. 

The collider phenomenology of the exotic fermions $\tilde{q}_3^B$ crucially depends on their masses. If the vector-like masses of the $\tilde{q}_3^A$, $\tilde{M}_A$, and of the $\tilde{q}_3^B$, $\tilde{M}_B$, are linked by the $Z_2$ symmetry, then theoretical and experimental constraints on the exotic quarks require $\tilde{M}_B \gtrsim 1$ TeV. In this case the exotic fermions have very suppressed electroweak pair production, but they can still be produced with significant rate in the decays of $\mathcal{X}$. We found that the ensuing signals are qualitatively similar to those of the exotic quarks, consisting primarily in $t\bar{t}$+MET. A potential handle to pin down the $\tilde{q}_3^B$ is to require an additional $Z$ boson, which is rarely produced in the decays of the $\tilde{q}_3^A$.

However, since the direct experimental constraints on the masses of the $\tilde{q}_3^B$ are rather weak, because they only carry SM electroweak charges, it is interesting to consider the case where $\tilde{M}_B \ll \tilde{M}_A\sim$ TeV. This soft $Z_2$ breaking does not generate 1-loop quadratic divergences for the Higgs mass, and hence does not introduce significant fine-tuning. The second part of the paper is therefore devoted to the study of this scenario. We found that in this case, at least one of the exotic fermions has a very suppressed decay width. As a consequence, its pair production through the Drell-Yan process is followed by the formation of twin QCD bound states, which then annihilate primarily into SM particles. The associated resonance signals, typical of Hidden Valley models \cite{Strassler:2006im}, provide a novel aspect in the phenomenology of the TH. We also discussed the significant effects that a relatively light $SU(6)$ exotic vector can have on the bound state phenomenology. Notice that while in this paper we focused on the decays of the bound states into SM particles, which are dominant, mixed SM/twin decays may provide rare but striking signatures if some of the twin particles are long-lived. For example, the charged bound state $\Upsilon_{\pm 0}$ can decay into $W^{\pm} \hat{g}\hat{g}$. If the hadronization of the twin gluons produces the lightest twin glueball, the latter can travel a macroscopic distance and decay through mixing with the Higgs, leading to the $W^\pm\,$+$\,(b\bar{b}$ displaced vertex) signature. 

We also sketched the phenomenology of the $SU(4)$ exotic vectors. Some of them have the right quantum numbers to mix with the SM $W$ and $Z$, and can therefore be singly produced in hadron collisions. We found that their contribution to the $S$ parameter of EWPT likely puts them out of the LHC reach, but they could be discovered at a future $100$ TeV collider. To avoid the large $T$ parameter constraint that would further push up their masses, the $SU(4)$ group may be extended to an $SO(8)$, which contains the custodial symmetry to protect the $T$ parameter. In this case there will be more exotic states, both in the vector and the fermion sectors. However, as shown in App.~\ref{SO8exo}, their phenomenologies are not qualitatively different and hence are well covered in this study.

\vspace{1cm}
\noindent{\bf Acknowledgments}

We thank Y.~Bai, A.~Coccaro, R.~Harnik, G.~Marques Tavares, T.~Tait and J.~Zupan for useful discussions, and D.~Redigolo for comments about the manuscript. This work was performed in part at the Aspen Center for Physics, which is supported by National Science Foundation grant PHY-1066293. HC and ES were supported in part by the US Department of Energy grant DE-SC-000999. YT was supported in part by the National Science Foundation under grant PHY-1315155, and by the Maryland Center for Fundamental Physics.

\appendix
\section{Two-site model}\label{SU4exo}
We describe here a two-site model where the gauge symmetry is $SU(6)\times SU(4)\times U(1)_X$ on the first (`strong') site, and $SU(3)^2 \times [SU(2)\times U(1)]^2 \subset SU(6)' \times [SU(4)'\times U(1)_X']$ on the second (`elementary') site. We begin by describing the electroweak sector, where the gauge symmetries are broken to the diagonal subgroup by the VEV $\left\langle \Sigma \right\rangle = f_d \mathbf{1}_4$ of a link field $\Sigma$ that transforms as $\Sigma \to U \Sigma V^\dagger$, where $U\in SU(4)$ and $V \in SU(4)'$. The covariant derivative for $\Sigma$ is then
\begin{equation}
D_\mu \Sigma = \partial_\mu \Sigma - i (T^a W_{s\,\mu}^a) \Sigma + i \Sigma (T^{\prime\,a} W_{el\,\mu}^a).
\end{equation}
The strong gauge fields are defined as
\begin{equation} \label{strong fields}
(T^a W_{s\,\mu}^a) = g_{\rho} \begin{pmatrix} \frac{\sigma^a \tilde{W}_B^a}{2} & \frac{\mathcal{W}}{\sqrt{2}} \\ \frac{\mathcal{W}^\dagger}{\sqrt{2}} & \frac{\sigma^a \tilde{W}_A^a}{2} \end{pmatrix} + g_\rho T_d^{SU(4)} \tilde{S} + g_X X_\Sigma \sqrt{2}\, \mathbf{1}_4 \tilde{X}\,,
\end{equation}
where the $2\times 2$ matrix $\mathcal{W}$ was defined in Table~\ref{Tab:charges} together with the quantum numbers of its components, and $T_d^{SU(4)} = \mathrm{diag}(\mathbf{1}_2, - \mathbf{1}_2)/(2\sqrt{2})$. Setting $g_X = g_\rho$ and $X_{\Sigma} = -1/4$, Eq.~\eqref{strong fields} can be rewritten as
\begin{equation} \label{cmp fields}
(T^a W_{s\,\mu}^a) = g_{\rho} \begin{pmatrix} \frac{\sigma^a \tilde{W}_B^a}{2} & \frac{\mathcal{W}}{\sqrt{2}} \\ \frac{\mathcal{W}^\dagger}{\sqrt{2}} & \frac{\sigma^a \tilde{W}_A^a}{2} \end{pmatrix} - g_\rho \begin{pmatrix} \frac{\tilde{B}_B}{2} & \\ & \frac{\tilde{B}_A}{2} \end{pmatrix}
\end{equation}
where $\tilde{B}_{B,A} \equiv (\mp \tilde{S} + \tilde{X})/\sqrt{2}$. It is immediate to see that the SM and twin hypercharges are identified with $Y = X + T_d/\sqrt{2}$ and $D = X - T_d/\sqrt{2}$ (for fields neutral under $SU(6)$), matching the normalization of Ref.~\cite{Cheng:2015buv}. The elementary gauge bosons read instead
\begin{equation}
(T^{\prime\,a} W_{el\,\mu}^a) = \begin{pmatrix} g_2\frac{\sigma^a W_B^a}{2} - g_1 \frac{B_B}{2} & \\ & g_2\frac{\sigma^a W_A^a}{2} - g_1 \frac{B_A}{2} \end{pmatrix} .
\end{equation}
The kinetic term of $\Sigma$ is $\mathcal{L}_{\Sigma} = \mathrm{Tr}[(D_\mu\Sigma)^\dagger (D^\mu \Sigma)]/4$, which setting $\Sigma$ to its VEV yields
\begin{align}
\mathcal{L}_\Sigma =&\, \frac{g_\rho^2 f_d^2}{4}\sum_{i=1}^2 (\mathcal{W}_i^{0} \mathcal{W}_i^{0\ast} + \mathcal{W}_i^{+} \mathcal{W}_i^-) + \frac{f_d^2}{8}\Big[(g_\rho^2 \tilde{W}_{A}^{3\,2} + g_\rho^2 \tilde{B}_A^2 + 2 g_\rho^2 \tilde{W}_A^+ \tilde{W}_A^-) + (g_2^2 W_{A}^{3\,2} + g_1^2 B_A^2 \nonumber \\ +&\, 2 g_2^2 W_A^+ W_A^-) 
-2 (g_\rho g_2 \tilde{W}_A^+ W_A^- + \mathrm{h.c.} + g_\rho g_2 \tilde{W}_A^3 W_A^3 + g_\rho g_1 \tilde{B}_A B_A) \Big] + (A \to B).
\end{align}
Thus the $\mathcal{W}$ acquire a mass $g_\rho f_d / 2$, while the mixing between strong and elementary states is diagonalized by, for example for the charged `$A$' states:
\begin{equation} \label{cmpelemmixing}
\begin{pmatrix} \tilde{W}_A^- \\ W_A^- \end{pmatrix} \to \frac{1}{\sqrt{g_\rho^2 + g_2^2}} \begin{pmatrix} g_\rho & g_2 \\ - g_2 & g_\rho \end{pmatrix} \begin{pmatrix} W_A^{'\,-} \\ W_A^{\rm SM\,-} \end{pmatrix}
\end{equation}
which leaves the $W_A^{\rm SM\,-}$ massless, while $m_{W'_A}^2 = (g_\rho^2 + g_2^2)f_d^2/4$. Similar rotations diagonalize the mixing of the neutral `$A$' and all the `$B$' fields. 

We also need to take into account the symmetry breaking due to the VEV of $H$, which lives in the strong site and also has $X_H = - 1/4$. Its covariant derivative is $D_\mu H = \partial_\mu H - i (T^a W_{s\,\mu}^a) H$, with $(T^a W_{s\,\mu}^a)$ defined in Eq.~\eqref{cmp fields}. In the vacuum of Eq.~\eqref{H UG}, the kinetic term $\mathcal{L}_H = (D_\mu H)^\dagger D^\mu H$ gives
\begin{align} \label{LH}
\mathcal{L}_H = \frac{g_\rho^2 f^2}{4} \Big\{&\,c_h^2 (\mathcal{W}_1^0 \mathcal{W}_1^{0\ast} + \mathcal{W}_1^+ \mathcal{W}_1^-) + s_h^2 (\mathcal{W}_1^0 \mathcal{W}_1^{0\ast} + \mathcal{W}_2^0 \mathcal{W}_2^{0\ast}) \nonumber \\ 
+&\, \tfrac{1}{2}c_h^2 [(\tilde{W}_B^3 - \tilde{B}_B)^2 + 2 \tilde{W}_B^+ \tilde{W}_B^-] + \tfrac{1}{2} s_h^2 [(\tilde{W}_A^3 - \tilde{B}_A)^2 + 2 \tilde{W}_A^+ \tilde{W}_A^-] \nonumber \\
+&\, s_h c_h [(\tilde{W}_B^3 - \tilde{B}_B)\mathcal{W}_{1R}^0 + \tilde{W}^+_B \mathcal{W}_2^0 + \mathrm{h.c.}] \nonumber \\
+&\, s_h c_h [(\tilde{W}_A^3 - \tilde{B}_A)\mathcal{W}_{1R}^0 + \tilde{W}^-_A \mathcal{W}_1^+ + \mathrm{h.c.}]\, \Big\}
\end{align}
where we have defined $\mathcal{W}_{1R}^0 \equiv (\mathcal{W}_1^0 + \mathrm{h.c.})/\sqrt{2}$. Notice that the imaginary counterpart of $\mathcal{W}_{1R}^0$ does not mix with any other field. The masses of the $\mathcal{W}$ resulting from $\mathcal{L}_\Sigma + \mathcal{L}_H$ were given in Eq.~\eqref{exomasses}, neglecting small corrections due to EWSB (i.e. proportional to $s_h$). At $O(s_h^2)$, the $W$ and $Z$ masses have the standard expressions $m_{W(0)}^2 = g^2v^2/4$ and $m_{Z(0)}^2 = (g^2 + g^{\prime\,2})v^2/4\,$, provided we identify the SM parameters as
\begin{equation} \label{SMparams}
g \equiv \frac{g_\rho g_2}{\sqrt{g_\rho^2 + g_2^2}}\,,\qquad g' \equiv \frac{g_\rho g_1}{\sqrt{g_\rho^2 + g_1^2}}\,,\qquad v \equiv \frac{f f_d}{\sqrt{f^2 + f_d^2}}s_h \simeq 246\;\mathrm{GeV}.
\end{equation}
The contribution to the $S$ parameter is computed from \cite{Barbieri:2004qk}
\begin{equation}
\hat{S} = \frac{g}{g'}\Pi'_{3B}(0)\,,
\end{equation}
where $\Pi_{3B}$ is the self-energy between the states $W^3_{A}$ and $B_A$, defined through the following effective Lagrangian in momentum space (the $W_A^\pm$ is omitted for brevity)
\begin{equation}
-\mathcal{L}_{\rm eff} = \frac{1}{2}\Pi_{33} W_{A}^3 W_{A}^3 + \frac{1}{2}\Pi_{BB} B_{A} B_{A} + \Pi_{3B}W_A^3 B_A\,,\qquad \Pi_{ij} = \Pi_{ij}(0) + p^2 \Pi'_{ij}(0) + \ldots\,. 
\end{equation}
Notice that it is natural to assume that the light SM fermions live on the elementary site. Therefore the strong (tilded) fields do not couple to light fermions, and once they are integrated out, only oblique corrections are generated \cite{Barbieri:2004qk}. To integrate them out we proceed diagrammatically, noticing that at $O(s_h^2)$ only exchange of the states $\tilde{W}_A^3, \tilde{B}_A$ and $\mathcal{W}_{1R}^0$ is relevant. The result was given in Eq.~\eqref{Sformula}. 

The $T$ parameter can be computed from the corrections to the $W,Z$ masses at $O(s_h^4)$, $m_{W(0)}^2 + \delta m_W^2$ and $m_{Z(0)}^2 + \delta m_Z^2$. For $g_{1,2}^2/g_\rho^2 \ll 1$ we find
\begin{equation}
\delta m_W^2 \simeq \frac{g^2}{g_\rho^2}\frac{g^2}{4}\frac{f^4 f_d^2 (f^2 + 2f_d^2) s_h^4}{(f^2 + f_d^2)^3}\,,\qquad  \delta m_Z^2 \simeq -\, \frac{(g^2+g^{\prime\,2})}{4} \frac{f^4 f_d^2 s_h^4}{(f^2 + f_d^2)^2}\,,
\end{equation}
leading to
\begin{equation}
\hat{T} = \frac{\delta m_W^2}{m_{W(0)}^2} - \frac{\delta m_Z^2}{m_{Z(0)}^2} \simeq \frac{v^2}{f_d^2}\,,
\end{equation}
where in the last equality we took the leading order in $g^2/g_\rho^2 \ll 1$, and employed the definition of $v$ in Eq.~\eqref{SMparams}. Quantitatively, requiring $\hat{T} < 10^{-3}$ gives $f_d \gtrsim 7.8$ TeV. 

In the color sector, the link field $\Sigma_6$ transforms as $\Sigma_6 \to U \Sigma_6 V^\dagger$, where $U\in SU(6)$ and $V \in SU(6)'$, and its VEV $\left\langle \Sigma_6 \right\rangle = f_c \mathbf{1}_6$ breaks the gauge symmetries to the diagonal subgroup. Its kinetic term reads $\mathrm{Tr}[(D_\mu\Sigma_6)^\dagger (D^\mu \Sigma_6)]/4$, where
\begin{equation}
D_\mu \Sigma_6 = \partial_\mu \Sigma_6 - i g_{\mathcal{X}} \begin{pmatrix} \frac{\lambda^i \tilde{G}_{B\mu}^i}{2} & \frac{\mathcal{X}_\mu}{\sqrt{2}} \\ \frac{\mathcal{X}_\mu^\dagger}{\sqrt{2}} & \frac{\lambda^i \tilde{G}_{A\mu}^i}{2} \end{pmatrix}  \Sigma_6 + i g_s \Sigma_6 \begin{pmatrix} \frac{\lambda^i G_{B\mu}^i}{2} &  \\  & \frac{\lambda^i G_{A\mu}^i}{2} \end{pmatrix},
\end{equation}
with $\lambda^i$ denoting the Gell-Mann matrices. Notice that in writing the fields on the strong site, we have neglected a singlet that is also contained in the adjoint of $SU(6)$. In the vacuum, the exotic vectors $\mathcal{X}$ acquire a mass $m_\mathcal{X} = g_\mathcal{X} f_c / 2$, whereas the mass mixing between the $\tilde{G}_A$ and $G_A$ is diagonalized by the rotation
\begin{equation} \label{cmpelemmixingstrong}
\begin{pmatrix} \tilde{G}_A \\ G_A \end{pmatrix} \to \frac{1}{\sqrt{g_{\mathcal{X}}^2 + g_s^2}} \begin{pmatrix} g_{\mathcal{X}} & g_s \\ - g_s & g_{\mathcal{X}} \end{pmatrix} \begin{pmatrix} \mathcal{G} \\ g \end{pmatrix},
\end{equation}
after which the SM gluon $g$ remains massless, while the mass of the KK gluon $\mathcal{G}$ is $m_{\mathcal{G}} = \sqrt{g_{\mathcal{X}}^2 + g_s^2} f_c/2$. The `$B$' fields are diagonalized in the same way.

The masses of the exotic fermions in Eq.~\eqref{eq:mass} arise from the following non-renormalizable operators
\begin{equation} \label{exomasses_operator}
- \mathcal{L}_m = \frac{1}{F} (y_A \bar{\xi}_A^{\,Ii} + y_B \bar{\xi}_B^{\,Ii})\, \Sigma^{\dagger ij}\, \Sigma_6^{\dagger IJ} Q_{3L}^{jJ} + \mathrm{h.c.},
\end{equation}
where $y_{A,B}$ are couplings and $F$ is a scale. For convenience, we have written Eq.~\eqref{exomasses_operator} as formally respecting a full $SU(6)\times SU(4)$ symmetry on each site, by introducing the spurions $\bar{\xi}_A^{\,Ii} \equiv \bar{\tilde{q}}^A_{3R}\, \delta^{IA} \delta^{iB}$ and $\bar{\xi}_B^{\,Ii} \equiv \bar{\tilde{q}}^B_{3R}\, \delta^{IB} \delta^{iA}$ on the elementary site. The indices $i,j\in\{B,A\}$ run over the two $SU(2)$ subgroups of $SU(4)$, while $I,J\in\{B,A\}$ correspond to the two $SU(3)$ subgroups of $SU(6)$. Setting the link fields to their diagonal VEVs, we obtain $\tilde{M}_{A,B} = y_{A,B} f_d f_c / F\,$.

\section{Extension to $SO(8)$}\label{SO8exo}
In this appendix we comment about the additional states that are present in TH models where the global symmetry is extended from $SU(4)$ to $SO(8)$. In the gauge sector, the adjoint representation of $SO(8)$ decomposes under the $SO(4)_A\times SO(4)_B$ subgroup as $\,\mathbf{28}\sim (\mathbf{6},\mathbf{1}) + (\mathbf{1},\mathbf{6}) + (\mathbf{4},\mathbf{4})$. The $(\mathbf{6},\mathbf{1})$ contains $(W_{AL}^a, W_{AR}^a)$ ($a=1,2,3$), where the $W_{AL}^a$ are the gauge bosons of $SU(2)_A^L$, identified with the $\tilde{W}_A^a$ of Eq.~\eqref{strong fields}, while the $W_{AR}^a$ are new vectors that gauge $SU(2)_{A}^R$. The phenomenology of the latter has been studied in the context of custodial composite Higgs models, see for example Ref.~\cite{Contino:2006nn}. The decomposition of the $(\mathbf{1},\mathbf{6})$ is entirely analogous, with $A\to B$. The $(\mathbf{4},\mathbf{4})$ contains the exotic vectors, that include both the $\mathcal{W}$ and additional states. However, there will still be only one linear combination of the exotic vectors that mixes with the SM $W$ and $Z$, and can therefore be singly produced at hadron colliders. Its phenomenology will be qualitatively similar to the one we outlined for the $\mathcal{W}$.  

In the fermion sector, the number of fields in $Q_{3L}$ needs to be doubled to form a fundamental representation of $SO(8)$. In our formalism, this can be done by extending the fields carrying twin color as follows
\begin{eqnarray}
&&q_{3L}^A \to \begin{pmatrix} q_{3L}^A & X_L \end{pmatrix},\qquad \tilde{q}_{3L}^A \to \begin{pmatrix} \tilde{q}_{3L}^A & \tilde{q}_{3L}^{\prime A}\end{pmatrix},  \nonumber \\
&&q_{3L}^B \to \begin{pmatrix} q_{3L}^B & \hat{X}_L \end{pmatrix},\qquad \tilde{q}_{3L}^B \to \begin{pmatrix} \tilde{q}_{3L}^B & \tilde{q}_{3L}^{\prime B}\end{pmatrix},  
\end{eqnarray}
where the fields in each set of parentheses transform in the same way under the $SU(3)_A \times SU(3)_B \times SU(2)_A \times SU(2)_B$ gauge group, but have different hypercharges or twin hypercharges. All fields except $q_{3L}^A$, $q_{3L}^B$ acquire vector-like masses with their vector-like partners, as in Eq.~(\ref{eq:mass}). The SM and twin electric charges of these fields are
\begin{equation}
\begin{pmatrix} q_{3L}^A & X_L \end{pmatrix} \sim \begin{pmatrix} \frac{2}{3} & \frac{5}{3} \\ -\frac{1}{3} & \frac{2}{3} \end{pmatrix}_{\rm SM} \oplus 
\begin{pmatrix} 0 & 0\\ 0 & 0 \end{pmatrix}_{\rm twin} , \quad
\begin{pmatrix} \tilde{q}_{3L}^A & \tilde{q}_{3L}^{\prime A}\end{pmatrix} \sim \begin{pmatrix} \frac{2}{3} & \frac{2}{3} \\ \frac{2}{3} & \frac{2}{3} \end{pmatrix}_{\rm SM} \oplus 
\begin{pmatrix} 0 & 1\\ -1 & 0 \end{pmatrix}_{\rm twin},
\end{equation}
with the SM and twin charges swapped for $(q_{3L}^B \; \hat{X}_L)$ and $(\tilde{q}_{3L}^B \; \tilde{q}_{3L}^{\prime B})$.

The phenomenology of the $X$ doublet is familiar from custodial composite Higgs models, see for example Ref.~\cite{DeSimone:2012fs} and references therein. For the extra copy of the exotic quarks, the upper component of $\tilde{q}_3^{\prime A}$ has twin electric charge $+1$ and behaves similarly to $\mathcal{B}$, whereas the lower component has zero twin electric charge and can mix with the SM top, in analogy with $\mathcal{T}$. Therefore the phenomenology of the additional exotic quarks is not expected to differ qualitatively from the one presented in Ref.~\cite{Cheng:2015buv}. Likewise, the upper component of $\tilde{q}_3^{\prime B}$ has SM electric charge $+1$ and behaves similarly to $\mathcal{K}^-$, whereas the lower component has zero SM electric charge and can mix with the twin top, in analogy with $\mathcal{K}^0$. Thus the phenomenology of the exotic fermions is qualitatively well captured by restricting the attention to the $\tilde{q}_3^B$, as we did in this paper. 

%%%%%%%%%%%%
%\bibliographystyle{utphys}
%\bibliography{./}

\end{document}